\documentclass[a4paper,11pt]{article}
\usepackage{jheppub} % for details on the use of the package, please
\usepackage[compat=1.1.0]{tikz-feynman} %Used for drawing Feynman diagrams
\tikzfeynmanset{/tikzfeynman/momentum/arrow shorten = 0.3}
\tikzfeynmanset{/tikzfeynman/warn luatex = false}
\usepackage{cancel} %Used for crossing terms out
\usepackage{sectsty} % Allows customizing section commands
\usepackage{graphicx}
\usepackage{thmtools}
\usepackage{jheppub}
\usepackage[T1]{fontenc} % if needed
\usepackage{amsmath,amsfonts,amsthm,amssymb} % Math packages
\usepackage{mathtools}
\usepackage{braket}
\usepackage{simplewick}
\usepackage{slashed}
\usepackage{nccmath}
\usepackage{subcaption}
\allowdisplaybreaks

\newcommand{\dd}{\partial}
\newcommand{\vdot}{\!\cdot\!}
\newcommand{\trace}{\mathrm{tr}}
\newcommand{\ftrace}{\mathrm{Tr}}

\DeclarePairedDelimiter{\floor}{\lfloor}{\rfloor}

\theoremstyle{definition}
\newtheorem{definition}{Definition}[section]

\theoremstyle{example}
\newtheorem{example}{Example}[section]

\title{\boldmath On the perturbative aspects of deformed Yang-Mills theory}

\author[a,1]{J. Lai}
\emailAdd{jlai@physics.utoronto.ca}
\affiliation[a]{Department of Physics, University of Toronto, Toronto, ON M5S 1A7, Canada}
\abstract{Centre-stabilised $SU(N)$ Yang-Mills theories on $\mathbb{R}^3 \times S^1$ are QCD-like theories that can be engineered to remain weakly-coupled at all energy scales by taking the $S^1$ circle length $L$ to be sufficiently small. In this regime, these theories admit effective long-distance descriptions as Abelian $U(1)^{N-1}$ gauge theories on $\mathbb{R}^3$, and semiclassics can be reliably employed to study non-perturbative phenomena such as colour confinement and the generation of mass gaps in an analytical setting. At the perturbative tree level, the long-distance effective theory contains $(N-1)$ free photons with identical gauge couplings $g^2_3 \equiv g^2/L$. Vacuum polarisation effects, from integrating out heavy charged fields, lift this degeneracy to give $\floor{\frac{N}{2}}$ distinct values: $g^2(\frac{2}{L})\lesssim g_{3,\ell}^2 L \lesssim g^2(\frac{2\pi}{NL}) $. In this work, we calculate these corrections to one-loop order in theories where the centre-symmetric vacuum is stabilised by $2\leq n_f \leq 5$ massive adjoint Weyl fermions with masses of order $m_\lambda \sim \frac{2\pi}{NL}$, (also known as "deformed Yang-Mills,") and show that our results agree with those found in previous studies in the $m_\lambda \to 0$ limit. Then, we show that our result has an intuitive interpretation as the running of the coupling in a "lattice momentum" in the context of the non-perturbative "emergent latticised fourth dimension" in the $N\to \infty$, fixed-$NL$ limit.}
\begin{document} 
\maketitle
\flushbottom
\section{Introduction}
\label{sec:intro}
Analytical methods to study the long-distance properties of four-dimensional asymptotically free non-Abelian gauge theories are few and far between; broadly speaking, it is a difficult problem to handle because the flow to strong coupling causes theoretical control over the system to be lost at lo- energy scales. 
While there are known models that are well-behaved enough to be studied analytically, (e.g., Seiberg-Witten theory \cite{seibergwitten},) these typically require special structures such as supersymmetry, or otherwise make use of gauge-gravity duality arguments and string-inspired tools (such as in Ref. \cite{adscft}).
\par
Over the past years, studies performed on "centre-stabilised" gauge theories on $\mathbb{R}^3\times S^1$ have been remarkably fruitful for providing insight into the non-perturbative dynamics of four-dimensional gauge theories. 
These models are distinguished from the few known analytically-calculable models in four dimensions by the fact that they can be engineered to remain weakly-coupled at all energy scales, so that a semiclassical expansion in terms of objects defined in the UV theory is reliable and self-consistent.
\par
The basic idea behind these models is as follows: by compactifying $\mathbb{R}^4$ to $\mathbb{R}^3\times S^1$, and "deforming" the pure Yang-Mills (YM) theory by adding a non-local and non-renormalisable potential to the Lagrangian, the well-known deconfining phase transition (cf. thermal Yang-Mills \cite{gpy},) at small circle lengths $L$ can be circumvented, and the theory remains in the colour-confining phase for all values of $L$. Adiabatic continuity to the full $\mathbb{R}^4$ theory of ultimate interest can therefore be argued on grounds that the theories share identical (non-spacetime) global symmetries for all $L \in [0, \infty]$. That is, they belong in the same "universality class" \cite{unsalyaffe08,unsalyaffe2010}.
\par
To be certain, the non-renormalisable "deformed" theory that we are describing can be viewed as a lattice theory with a fixed finite lattice spacing \cite{unsal07}. On the other hand, it is also possible to define a UV-complete continuum theory with the same desired properties by introducing $n_f$ $S^1$-periodic adjoint-representation fermion fields to the pure YM Lagrangian: The desired deformation potential is realised as the fermionic contribution to the dynamically-generated Gross-Pisarski-Yaffe (GPY) effective potential at energy scales below $\sim \frac{1}{L}$ \cite{unsalyaffe08,unsalshifman08,unsal09,unsalyaffe2010}. If the fermions are massless,\footnote{It should be noted that QCD(adj) with $n_f$ massless fermions in its spectrum has a global chiral symmetry not shared by the $\mathbb{R}^4$ pure Yang-Mills and is therefore not covered by the aforementioned "universality class argument."} this class of theories is referred to as QCD(adj) if $2 \leq n_f \leq 5$, and super Yang-Mills (SYM) if $n_f=1$. It is called "deformed Yang-Mills" (dYM), when the $2 \leq n_f \leq 5$ fermions are massive, or if the deformation potential is added "by hand," as in the lattice formulation.
\par
From the theorist's perspective, one of the most alluring features of these admittedly artificial setups is that they admit a "weak-coupling regime" at $\Lambda NL \ll 2\pi$, (where $\Lambda$ is the strong-coupling scale,) in which the gauge coupling $g^2$ (and more pertinently, $g^2N$,) remains small at all energy scales. Thus, in this regime, the semiclassical expansion over high-energy monopole-instanton configurations is trustworthy, and can be reliably employed to study the effects of the non-perturbative physics on the low energy theory. The result is a theoretical laboratory in which a wide variety of non-perturbative low-energy phenomena can be studied analytically: For example, colour confinement, the generation of a non-perturbative mass gap \cite{unsalyaffe08,unsal09}, a deconfining phase transition \cite{continuity,symallgroups}, and certain aspects of chiral symmetry breaking \cite{unsal07}. For this reason, $\Lambda NL \ll 2\pi$ is sometimes also called the "calculable regime" in the context of centre-stabilised $\mathbb{R}^3\times S^1$ theories. For comprehensive reviews, see Refs. \cite{poppitzreview,newmethods,semiclassical2012}.
\medskip
\par
At leading perturbative order and finite $N$, the IR effective theory of $SU(N)$ dYM and QCD(adj) in the calculable regime is sometimes described as being "rather boring" \cite{poppitzreview}, because the gauge sector describes $(N-1)$ free, massless photons in $\mathbb{R}^3$. When vacuum polarisation effects are accounted for, the photons acquire $\floor{\frac{N}{2}}$. These corrections have been calculated to one-loop order by various methods when the fermions are assumed to be massless: for $N=2, \,3$ QCD(adj) in Ref. \cite{renormalons}, for SYM (i.e., $n_f=1$) with arbitrary $N$ in Ref. \cite{symallgroups}, and in QCD(adj) with arbitrary $N$ in Ref. \cite{pelizzani}.
\par 
In this study, we derive a more general expression for these corrections that in particular covers the massive fermion case, for generic masses $m_1,...\, m_{n_f}$ such that the theory remains in the centre-symmetric and weak-coupling regime. The final result is contained in Equation \eqref{eq:kappaell}, and the bulk of our exposition explains how we arrive at this result. 
Our motivating aim is to confirm that the perturbative corrections to finite-$N$ $SU(N)$ dYM theory yield no unpleasant surprises even when the stabiliser fermions are assumed to be heavy. This is a very reasonable assumption to make, since ultimately we are interested in obtaining insight on pure Yang-Mills on $\mathbb{R}^4$, and a continuum QCD-like theory that continues smoothly to pure YM should not contain light adjoint fermions in its IR spectrum.
\par
Nevertheless, our results are not entirely devoid of novelty: Ref. \cite{emergentdimensions} showed that in the $N\to\infty$, $L\to 0$, fixed-$NL$ limit of SYM, an emergent latticised fourth dimension appears, emerging out of the space of fields --- even though we should expect that taking $L\to 0$ ought to result in a 3d theory. In particular, this emergent dimension exhibits $z=2$ Lifschitz scaling invariance in SYM. In other words, the action is quartic, rather than quadratic, in the momentum: $\sim |\partial_y^2\Phi|^2$, where $\partial_y$ is the partial derivative in the emergent latticised dimension.
Simply put, this is because in SYM, there is a discrete $\mathbb{Z}_N$ chiral symmetry (not to be confused with the $\mathbb{Z}_N$ centre symmetry,) that forbids monopole-instantons from contributing a bosonic potential of the form $\sim |\partial_y \Phi|^2$ in the semiclassical expansion. Such a potential is permitted, however, when the chiral symmetry is explicitly broken by a non-zero fermion mass, as is in the case we study here. We find a satisfying and intuitive interpretation of our massive correction in this emergent dimension as the flow towards \textit{strong} coupling for \textit{large} values of the "lattice momentum."
\smallskip
\par
The rest of this paper is structured as follows: Section \ref{sec:background} contains an overview of the essentials of dYM theory in an effort to make this paper more self-contained. For the benefit of the impatient reader, we have placed our main result, Equation \eqref{eq:kappaell} and its accompanying discussion, in Section \ref{sec:results}. A discussion of this result in the context of the emergent latticised dimension of Ref. \cite{emergentdimensions} is contained in Section \ref{sec:latticedim}. 
\par
Section \ref{sec:maincalculation} covers the derivation of Equation \eqref{eq:kappaell} in detail, starting from the very beginning with the UV dYM Lagrangian. Since this calculation is fairly long and convoluted, we briefly summarise what we have done at the end of subsections \ref{sec:setup} and \ref{sec:wilsonian} to help the reader keep track of our progress. The main "meat" of the calculation, and therefore of this paper, is mostly contained in Section \ref{sec:outline}, especially Section \ref{sec:qtherm}.
\par
In our calculation, we use the Mellin transform to rewrite certain infinite sums in a form that allows their asymptotic behaviour to be more easily seen. The details of this manipulation, which is mostly just complex analysis in one variable, is given in Appendix \ref{sec:mellin}.
\section{Background, results and discussion}
\label{sec:background}
\subsection{Review of dYM: I. Perturbative aspects}
Consider pure $SU(N)$ Yang-Mills theory on compactified $\mathbb{R}^3\times S^1$, where the $S^1$ is a circle of circumference $L$: 
\begin{equation}
    S_{\text{YM}}[\mathcal{A},\mathcal{F}]= \int_{\mathbb{R}^3\times S^1 } \frac{1}{2g^2} \trace (\mathcal{F}^2).
\end{equation}
This theory enjoys a global $\mathbb{Z}_N= Z(SU(N))$ centre symmetry, as it only contains fields transforming in the adjoint representation of the gauge group. The action of this symmetry may be thought of as a "gauge"\footnotemark{} transformation $g(x^\mu,x^4):\mathbb{R}^3\times S^1\to SU(N)$ that is periodic over the $S^1$ modulo a $\mathbb{Z}_N$ factor: 
\begin{equation}
    g(x^\mu, 0) = \omega g(x^\mu, L) \,,\qquad \omega\equiv e^{i 2\pi/N}.
\end{equation}
This acts on the fundamental representation Polyakov loop $\Omega$, the gauge holonomy along the $S^1$,
\footnotetext{Though, of course, the $g(x^\mu,x^4)$ so defined is not a true gauge transformation by any means. That is, it is not a transition function between local trivialisations of the principle bundle.}
\begin{equation}
    \Omega\equiv \mathcal{P} \exp \, i\int_{0}^L \! dx^4 \mathcal{A}_4 ,
\end{equation}
as\footnote{According to the modern viewpoint, this $\mathbb{Z}_N$ belongs to a class of "generalised" global symmetries, which act on operators with non-trivial spatial extent. In this context, Equation \eqref{eq:censym} \textit{defines} the symmetry \cite{gensym}}.
\begin{equation}
    \mathbb{Z}_N : \trace \Omega \to \omega \, \trace  \Omega, \label{eq:censym}
\end{equation}
where $\mathcal{A}_4$ is the $S^1$ part of the gauge field $\mathcal{A}$.
\par
At large $L$, the centre symmetry is unbroken. That is to say, $\langle \trace \Omega^n \rangle = 0$ in the ground state for all $n\neq 0 \mod N$. On the other hand, it is a well-known fact \cite{gpy} that in the small-$L$ limit, the theory undergoes a deconfining phase transition associated with the breaking of centre symmetry: In this regime, where perturbative analyses can be trusted because of asymptotic freedom, Ref. \cite{gpy} showed that the theory produces a (GPY) effective potential, $V_{\text{pert.}}[\Omega]$:
\begin{equation}
    V_{\text{pert.}}[\Omega] = -\frac{2}{\pi^2 L^4}\sum_{n=1}^{\infty}\frac{|\trace \Omega^n|^2}{n^4}(1 + O(g^2)).
    \label{eq:gpy}
\end{equation}
This result can be found by integrating out the Kaluza-Klein modes at one-loop order. This potential is minimised by $\Omega $ of the form $ \Omega = \omega^k\, 1_{N}$ for any integer $k$, suggesting that the theory has $N$ degenerate vacua related by the $\mathbb{Z}_N$ symmetry and describes a gluon plasma phase.
\par
The basic idea behind $\mathbb{R}^3\times S^1$ theories such as dYM is to re-enforce the stability of the $\mathbb{Z}_N$ at small $L$ by "flipping" the shape of the GPY potential, so to speak. This can be done in the most direct way by simply adding a "double-trace" term to the YM action:
\begin{subequations}
    \begin{equation}
        \mathcal{L}_{\text{dYM}}= \mathcal{L}_{\text{YM}} + V_{\text{deformed}}[\Omega] ,
    \end{equation}
where
    \begin{equation}
        V_{\text{deformed}}[\Omega] =  \frac{1}{L^4}\sum_{n=1}^{\floor{N/2}} a_n |\trace \Omega^n|^2 .
        \label{eq:deformed}
    \end{equation}
\end{subequations}
But such a term is manifestly non-local, being defined in terms of a non-local operator. It is also non-renormalisable, as it contains infinitely many irrelevant operators which blow up uncontrollably in the UV. As such, such a deformation of the theory may be considered problematic to those with a philosophical preference for continuum theories. We will return to address this objection later, and focus on the effects of the double-trace deformation potential on the IR theory for now.
\par
The $\mathbb{Z}_N$ symmetry is said to be preserved if and only if the vacuum state of the theory satisfies $\langle \trace \Omega^n \rangle = 0$ for all $n\neq 0 \mod N$, so the coefficients $a_n>0$ in Equation \eqref{eq:deformed} must each be chosen so as to dominate the dynamically generated $V_{\text{pert.}}[\Omega]$. With the centre symmetry stabilised, we can remove the gauge redundancy of $\Omega$ by choosing a diagonal representative from the class of physically equivalent minima:
\begin{equation}
    \Omega  = \omega^{(1-N)/2} \textrm{diag}(1,\omega,...,\omega^{N-1}).
    \label{eq:omegavev}
\end{equation}
This choice is in fact unique, up to permutations of the coefficients corresponding to Weyl reflections that can also be gauged-fixed away by working in the (affine) Weyl chamber. This allows us to write $\langle \Omega \rangle$ as a physically meaningful expectation value despite its uncontracted matrix indices.\footnote{To be certain, the action of the centre $\mathbb{Z}_N$ in this gauge is, with $\Omega_{i}$ denoting the $i$'th diagonal of $\Omega$ \cite{semiclassical2012},
\begin{equation}
    \mathbb{Z}_N : \Omega_{i} \to 
    \begin{cases} 
        \omega \, \Omega_{i-1} & i \neq 1 , \\
        \omega \, \Omega_{N} & i = 1 .
    \end{cases}
\end{equation}
The cyclic permutation is necessitated by gauge-fixing to the Weyl chamber.
}
\par
This vev precipitates a simulacrum of the Higgs mechanism in which the $\mathcal{A}_4$ field plays the role of an adjoint Higgs field. The gauge group generators left unbroken by $\langle \Omega \rangle$ form a Cartan subalgebra $\mathfrak{t} \subset su(N)$, generating the maximal torus $U(1)^{N-1} \subset SU(N)$.
The Higgs mechanism endows fields in $\mathfrak{t}^\perp$ (i.e., fields that carry charge under the $U(1)^{N-1}$) with an effective mass $\geq \frac{2\pi}{NL} \equiv m_W$, the so-called Abelianisation scale. 
\par
We can now perform the path integral around the centre symmetric vacuum: Working perturbatively, (the treatment of the non-perturbative physics is left to Section \ref{sec:nonpert},) weak-coupling ensures that the $\mathcal{A}_4$ fluctuations around $\langle \Omega \rangle$, of mass $ \gtrsim \sqrt{g^2 N} \, m_W$, can only effect small corrections to the effective action. Weak-coupling, in turn, is guaranteed by the weak-coupling assumption $m_W\gg\Lambda$ --- meaning that all dynamic charged fields can be safely integrated out before the onset of strong coupling as we carry the theory towards the infrared. When the dust settles, we are left with a weakly-coupled $U(1)^{N-1}$ gauge theory containing no light charged fields in its spectrum. In fewer words: everything works out fine.
\par
In settings where it is desirable to have a theory that respects both locality and UV-completeness, and yet preserve centre symmetry at all scales, we can opt to have $V_{\text{deformed}}[\Omega]$ generated dynamically as well, by adding sufficiently light, or massless, $S^1$-periodic adjoint fermions $\lambda_I$ to the theory \cite{unsal09}, rather than inserting the deformation potential "by hand." In such a setting, the periodicity requirement $\lambda_I(x^\mu , x^4 ) =+ \lambda_I(x^\mu , x^4+L )$ prohibits a thermal interpretation for the $S^1$, which must therefore be taken to be a spatial circle.
\par
For the theory with $1 \leq n_f \leq 5$ Weyl fields\footnote{The theory loses asymptotic freedom for $n_f>5$.} of masses $m_I$ indexed by $1\leq I\leq n_f$, the dynamically generated (GPY) effective potential is \cite{symallgroups,unsalyaffe2010,misumi2014}
\begin{equation}
       V[\Omega] = -\frac{1}{\pi^2 L^4}\sum_{n=1}^{\infty}\frac{1}{n^4}\Big [ 2 - \sum_{I=1}^{n_f}(nL m_I)^2K_2 (nL m_I)\Big ]|\trace \Omega^n|^2 ,
       \label{eq:potential}
\end{equation}
where $K_2$ is the modified Bessel function of order 2. It is not hard to find constraints on the $m_I$ for each value $n_f$ that stabilise the centre-symmetric $\Omega$ in Equation \eqref{eq:omegavev}; see e.g., Ref. \cite{misumi2014}. We also note in passing that the $n_f=1$ potential vanishes (in fact, to all perturbative orders,) in the massless case, and is centre-unstable otherwise. This particular case is known as super Yang-Mills (SYM), in which the UV theory enjoys an exact $\mathcal{N}=1$ supersymmetry, which allows many aspects of its rich non-perturbative physics to be calculated exactly. But SYM is outside of the scope of this study, along with the massless QCD(adj), and we henceforth only consider $2\leq n_f\leq 5$ and $m_I>0$. 
\par
Assuming the fermion masses to be roughly equal, it turns out that centre stability requires $m_I \lesssim m_W$. In particular, this means that we can assume that the $m_I$ are $O(m_W)$ so that the fermions disappear from the low-energy theory, and the effective action can be written on $\mathbb{R}^3$ as:
\begin{subequations}
    \begin{equation}
    S_{3d} = \int_{\mathbb{R}^3}\sum_{a,b=1}^N \kappa_{ab} F_{\mu\nu}^a F_{\mu\nu}^b  + (\text{$A_4$ and higher order terms})\label{eq:eftaction}
    \end{equation}
for Abelian field strengths $F_{\mu\nu}^a$ and $\mathbb{R}^3$ indices $\mu,\nu \in \{1,2,3\}$ and Lie algebra indices $a,b \in \{1\,...\,N\}$. There is also a a neutral scalar field $A_4^a$ in the IR theory, which descends from the Abelian part of $\mathcal{A}_4$ and corresponds to the oscillations of the eigenvalues of $\Omega$ around the centre-symmetric vev. But this field receives a (mass)$^2$ $\sim g^2 N m_W^2$ correction from the GPY potential and can be integrated out by moving the theory to still lower energies. 
\par
The quantity $\kappa_{ab}$ in Equation \eqref{eq:eftaction} is the quantum-corrected photon coupling matrix:
    \begin{equation}
        \kappa_{ab} = \frac{m_W^{-1}}{16\pi} \Big( \frac{8\pi^2}{Ng^2}\delta_{ab} +\,O(1)\, \Big ) \,,\qquad g^2\equiv g^2(4\pi/L) ,
        \label{eq:treekappa}
    \end{equation}
\end{subequations}
where in Equation \eqref{eq:treekappa}, the gauge coupling is normalised with respect to the $L\to\infty$, $m_I\to 0$ limit:
    \begin{equation}
        \Lambda^{b_0} = \mu^{b_0} \exp \Big (-\frac{8\pi^2}{g^2(\mu) N} \Big ) \,,\qquad b_0 \equiv \frac{11-2n_{f}}{3} ,
        \label{eq:lambdadef}
    \end{equation}
and $b_0$ is the one-loop coefficient of the beta function of $(g^2 N)^{-1}$ in that limit. As stated before, these corrections have been calculated in previous studies for arbitrary $N$ and $1 \leq n_f\leq5$ in the limit $m_I=0$. Our calculation generalises to the massive case, and is a new result. We also believe it to be a non-trivial problem in terms of significance (as we will argue in this Section,) as well difficulty (which we will demonstrate in the next).

\subsubsection{The one-loop corrections to $\kappa_{ab}$}
\label{sec:results}
\begin{figure}[t]
    \includegraphics[width=12cm]{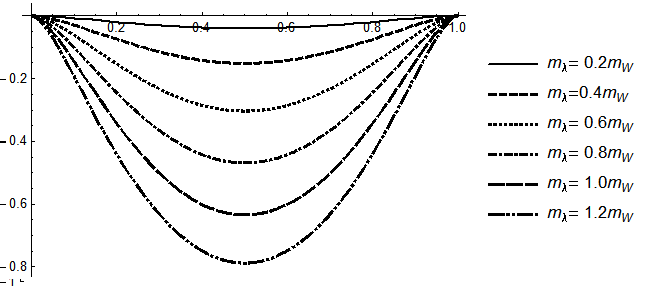}
    \centering
    \caption{A numerical plot of $W_\ell$ as a function of $\frac{\ell}{N}$ for some select values of $m$, in units of $m_W$. The graphs were plotted with Equation \eqref{eq:mediumWell} for $m=1.0 m_W$ and $m=1.2m_W$, and with \eqref{eq:well} otherwise. }
    \label{fig:well}
\end{figure}

Compared to the writing out the matrix entries of $\kappa_{ab}$ explicitly in the Cartan-Weyl basis, (given in Equation \eqref{eq:kab},) it is more enlightening to present its eigenvalues, $\kappa_\ell$:
\begin{subequations}
    \begin{equation}
        \sum_{a,b=1}^{N}\kappa_{ab} F_{\mu\nu}^a F^{b\mu\nu} = \sum_{\ell=1}^{N} \kappa_\ell \tilde{F}_{\mu\nu}^\ell \tilde{F}^{\ell\mu\nu},
    \end{equation}
    where
    \begin{equation}
        \tilde{F}_{\mu\nu}^\ell \equiv \frac{1}{\sqrt{N}}\sum_{a=1}^{N}  \omega^{-\ell a} F_{\mu\nu}^a\, , \qquad \kappa_\ell \equiv \frac{1}{N}\sum_{a,b=1}^{N}  \omega^{\ell(a-b)}\kappa_{ab} ,
    \end{equation}
    \label{fourier}
\end{subequations}
where again $\omega = e^{i\frac{2\pi}{N}}$, so Equation \eqref{fourier} are really just discrete Fourier transforms in the indices $a,b$.
%%emphasise this is main result below:
Then, assuming centre-stabilising fermion masses $m_I$,\footnote{Please note that the $\ell=N$ mode must be excluded from the spectrum as it corresponds to the trace of $F^a_{\mu\nu}$, (as can be seen from the definition,) which is unphysical in our theory; it would have been a physical mode if we had instead chosen our gauge group to be $U(N)$.}
\begin{equation}
     \begin{split}
         \kappa_{\ell} &\equiv\frac{1}{4 } g_{3,\ell}^{-2} \\
         &= \frac{m_W^{-1}}{16\pi} \Bigg [ \frac{8\pi^2}{N g^2(m_W e^{-\gamma})} + \Big ( \frac{11-2n_f}{3} \Big ) \log \frac{1}{\sin \pi \frac{\ell}{N}} + \frac{2}{3} \sum_{I=1}^{n_f} W_\ell \Big ( \frac{m_I}{m_W} \Big )  \Bigg ] \\
         & \qquad \qquad \qquad \qquad\qquad \qquad \qquad \qquad\qquad \qquad \qquad   \text{for}\,\,1\leq\ell\leq N-1 .
    \label{eq:kappaell}
     \end{split}
\end{equation}
$W_\ell=W_{N-\ell}$ is an $O(1)$ pure function of the masses $m_I$ in units of $m_W$, which enjoys the following properties:
\begin{subequations}
    \begin{align}
        W_\ell(0) = 0& \qquad \text{for all integers $\ell$,} \label{eq:masslessW}\\
         W_\ell \Big ( \frac{m}{m_W} \Big ) \to \log \Big ( \frac{m_W e^{-\gamma}}{m \sin \pi \frac{\ell}{N}}   \Big )& \qquad  \text{monotonically, as}\,\,\, m/m_W \to \infty, \label{eq:asympW} \\
         W_\ell \Big ( \frac{m}{m_W} \Big ) \to 0& \qquad  \text{as}\,\,\,\frac{\ell}{N} \to 0\,,\, 1\,,\\
         W_\ell (\tau ')< W_\ell (\tau) < 0& \qquad \text{for}\,\,\, 0<\tau<\tau', \label{eq:tracemode}\\
         W_{\floor{N/2}} (\tau) < W_{\ell'} (\tau)< W_{\ell} (\tau) \leq 0& \qquad \text{for}\,\,\,  \ell<\ell'< \floor{N/2}\, ,\quad \tau>0 . \label{eq:ineqe} 
    \end{align}
    \label{eq:wellproperties}
\end{subequations}
A plot of $W_\ell$ as a function of $\frac{\ell}{N}$ for a few select values of $m$ is given in Figure \ref{fig:well}.
\par
Equation \eqref{eq:kappaell} is written so that all the information about the one-loop corrections due to the fermion masses is encoded in the pure function $W_\ell$. In particular, Equation \eqref{eq:tracemode} implies that the $\ell=1$ mode receives a vanishingly small mass correction in the $N\to\infty$ limit; conversely, Equation \eqref{eq:ineqe} implies the $\floor{N/2}$ mode receives the largest mass correction. Equations \eqref{eq:masslessW} and \eqref{eq:asympW} together imply that $\kappa_{1} \geq \kappa_\ell \geq \kappa_{\floor{N/2}}$ for all $m_I$ and $\ell$. Note that in order for our results to make sense, we must require $\kappa_{\floor{N/2}}>0$; we will discuss the conditions that fulfill this requirement later.
\par
We present two analytic expressions for $W_\ell$ with different convergence properties: one expression holds for $m< m_W$, and the other, for $m\gtrsim m_W$. The former of these is:
\begin{subequations}
    \begin{equation}
        \begin{split}
              W_\ell \Big ( \frac{m}{m_W} \Big )  &= \sum^\infty_{n=1} \frac{(-1)^n(2n)!}{ 2^{2n} (n!)^2 }\Big (\frac{m}{m_W} \Big)^{2n}\Big [ \zeta(2n+1) - Re \Big ( \mathrm{Li}_{2n+1}( e^{2\pi i \frac{\ell}{N}})  \Big ) \Big ] \\
              & \qquad\qquad\qquad\qquad\qquad\qquad\qquad\qquad \qquad \qquad \text{for} \,\, (m/m_W) <1 ,
        \end{split}
        \label{eq:well}
    \end{equation}
where $\zeta$ is the Riemann zeta function, and $\mathrm{Li}_{s}$ is the polylogarithm function of order $s$. This expression fails to converge when $m>m_W$\footnotemark; it is in this regime where our second expression is more useful:
    \begin{equation}
       \begin{split}
           W_\ell \Big ( \frac{m}{m_W} \Big )  = & \log \Big[ \frac{m_W e^{-\gamma}}{m \sin \pi \frac{\ell}{N}}   \Big ) \\ 
           & \quad + \sum_{p=1}^\infty \Big \{ 2 K_0\Big ( 2\pi p \frac{m}{m_W}\Big ) - K_0\Big ( 2\pi \Big ( p-1 + \frac{\ell}{N}  \Big )\frac{m}{m_W}\Big )  \\
           &\qquad\qquad - K_0 \Big ( 2\pi \Big (p- \frac{\ell}{N}  \Big ) \frac{m}{m_W}\Big ] \Big \} .
        \label{eq:mediumWell}
       \end{split}
    \end{equation}
\end{subequations}
$K_0$ is the modified Bessel function of order $0$. Equation \eqref{eq:mediumWell} is one-loop exact for all $m$, but it is more useful at large $m\gtrsim m_W$, where it may be very well approximated by the first term of the series, as $K_0(t) \sim e^{-t}$ at large $t$. Conversely, Equation \eqref{eq:mediumWell} is less useful at small $m/m_W$ as $K_0(t) \sim \log t$ at small $t$. 
\footnotetext{Observing that the terms in the square brackets of \eqref{eq:well} are absolutely bounded for all $n\geq 1$, and that
\begin{align*}
    \frac{(2n)!}{(n!)^2 2^{2n}} = \frac{1}{\sqrt{\pi}}\frac{\Gamma(n+\frac{1}{2})}{\Gamma(n)} ,
\end{align*}
the root test gives a radius of convergence of $(m/m_W)<1$.\label{fn:roottest}}
\par
Since $W_\ell\leq 0$, the massive correction competes against the massless-limit corrections encoded in the log-sine term. Indeed, by taking $m_I\gg m_W$, the $I$'th fermion decouples from the theory\footnotemark{}, $n_f\to (n_f-1)$, up to an overall renormalisation of $g^2N$, or, equivalently, a re-definition of the strong-coupling scale $\Lambda$.
\footnotetext{Assuming, of course, that the theory still remains in the centre-symmetric regime.}
\par
\medskip
We can also use Equation \eqref{eq:lambdadef} to define a "lattice-renormalised" 't Hooft coupling $\lambda_\ell$:
\begin{subequations}
    \begin{equation}
      \begin{split}
          \frac{1}{\lambda_\ell} \equiv  b_0 \log \Big ( \frac{m_W e^{-\gamma}}{\Lambda \sin \pi \frac{\ell}{N}}  \Big )\, \quad  \text{for}\,\,1\leq\ell\leq N-1 .
    \label{eq:latticecoupling}
      \end{split}
\end{equation}
Then assuming for convenience equal fermion masses $m_I = m_\lambda$ for all $I=1\,...\,n_f$, and abbreviating $ W_\ell(\frac{m_\lambda}{m_W}) = W_\ell$, Equation \eqref{eq:kappaell} can be written in a neater form:
\begin{equation}
    \kappa_\ell = \frac{m_W^{-1}}{16\pi}\Big(\frac{1}{\lambda_\ell} + \frac{2n_f }{3} W_\ell\Big) \quad  \text{for}\,\,1\leq\ell\leq N-1 .
\end{equation}
\end{subequations}
\begin{figure}[t]
    \includegraphics[width=12cm]{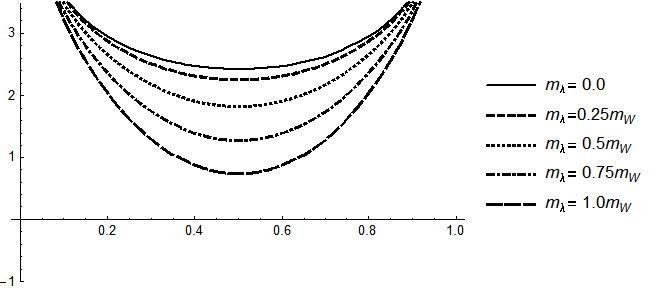}
    \centering
    \caption{A numerical plot of $\kappa_\ell$ in units of $\frac{m_W^{-1}}{16\pi}$, as a function of $\frac{\ell}{N}$ for some select values of $m_\lambda$, for $n_f=4$ and $m_W= e^3 \Lambda$. The plot diverges at $\ell/N \to 0, 1$ (not depicted) due to the log-sine running in $\frac{1}{\lambda_\ell}$.}
    \label{fig:kell}
\end{figure}
The dependence of $\kappa_\ell$ on $\frac{\ell}{N}$ is illustrated in Fig. \ref{fig:kell} for a few sample values of $m_\lambda$, for fixed $nf=4$ and $m_W=e^3\Lambda$. 
Given a fixed value of $m_W/\Lambda= \frac{2\pi}{\Lambda N L} \gg 1$, it is a straightforward exercise in numerical analysis to find constraints on $m_\lambda$ so that $\kappa_{\floor{N/2}} > 0$ in order for our result to make sense. Conversely, $\mathbb{Z}_N$-stability requires that e.g. $m_\lambda\lessapprox1.08m_W$ for $n_f=4$, and this bound can in particular be saturated by taking $m_W/\Lambda \gtrapprox e^3$. On the other hand, $\mathbb{Z}_N$ stability requires $m_\lambda\lessapprox1.2m_W$ for $n_f=5$, and saturation of that bound would require $m_W/\Lambda \gtrapprox e^{9}$, which is substantially larger.
\par
\medskip
Equations \eqref{eq:kappaell}, \eqref{eq:well}, and  \eqref{eq:mediumWell} comprise the main results of this paper; they are derived in detail in Section \ref{sec:maincalculation}, with reference to some results from Appendix \ref{sec:mellin}. The rest of this Section discusses how to interpret these results in the context of the non-perturbative physics of dYM theory, particularly with regards to the "emergent latticised dimension" of Ref. \cite{emergentdimensions}.
\par
\subsection{Review of dYM: II. Non-perturbative aspects}
\label{sec:nonpert}
Let us now very quickly summarise the derivation of the low-energy effective Lagrangian in dYM theory at leading order in the semiclassical expansion. The basic idea is essentially the same as Polyakov's version of confinement in the Georgi-Glashow (GG) model in $2+1$ dimensions \cite{shifman2012advanced}, although there are crucial differences due to the intrinsically four-dimensional nature of dYM theory.
The reader interested in a more detailed exposition is referred to Refs. \cite{unsalyaffe08,stringtensions,smallqcd}. 
\par
The contribution of the non-perturbative physics to the path integral in a weakly-coupled Euclidean QFT can be approximated to first exponential order by summing over classical field configurations that are inundated by a "gas" of weakly interacting minimal-action instantons. This is the so-called dilute instanton gas approximation, and it is applicable in dYM because weak coupling can be reliably assumed to hold at all scales provided that $N L \Lambda \ll 2\pi$.
\par
In addition to a topological charge $Q \sim \int \trace \, \mathcal{F} \wedge \mathcal{F} = \frac{1}{N}$, the instantons of dYM theory carry a magnetic charge ( $\sim \int F$) under the $U(1)^{N-1}$ --- they are essentially 't Hooft-Polyakov monopoles, with $\mathcal{A}_4$ again standing in for the adjoint Higgs field. In particular, we call them monopole-instantons.
\par
Among these, there are $(N-1)$ "BPS"\footnotemark monopoles, each carrying a magnetic charge corresponding to a simple root $\alpha_i$ of the gauge group.
\footnotetext{This is a common abuse of terminology: outside of the supersymmetric case, the BPS bound cannot be saturated because the "Higgs" potential $V[\Omega]$ cannot be set to zero. So strictly speaking we are expanding around "almost-BPS" configurations.}
In distinction to the 3-dimensional Polyakov model, in $\mathbb{R}^3\times S^1$ theories there is also an $N$'th "twisted," or Kaluza-Klein, (KK) monopole, which carries charge $\sum_{i=1}^{N-1} (- \alpha_i) \equiv \alpha_N$, the affine root. In addition to these, there are also the antiparticles carrying charge $-\alpha_i$. 
In a sense, the ($N-1$) BPS+KK monopole-instantons can be thought of as the "dissociation" of the BPST instanton in 4-dimensional $SU(N)$ Yang-Mills into $N$ sub-constituents \cite{lee1997,kraan1998}.
\par
We unfortunately do not have exact expressions for the monopole-instantons outside of the supersymmetric $n_f=1, m=0$ case. But as it turns out, they will not be required as far as our presentation is concerned: we need only know that these charged objects interact with a long-range Coulombic interaction, and have a non-linear "$A_4$/Higgs condensate" core of size $\sim m_W^{-1}$. In addition, there is also a "medium-range" ($\sim 1/g\sqrt{N} m_W$) Yukawa interaction arising from $A_4$/Higgs exchange. 
\par
\begin{figure}[t]
    \includegraphics[width=12cm]{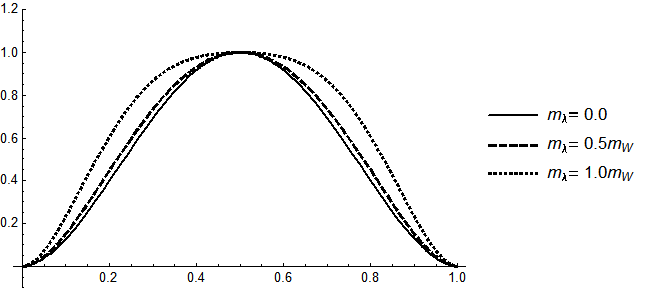}
    \centering
    \caption{A numerical plot of $m_{\sigma,\ell}^2/m_{\sigma,N/2}^2$ as a function of $\frac{\ell}{N}$, for $n_f=4$ and $m_W= e^4 \Lambda$. }
    \label{fig:mell}
\end{figure}

Every insertion of a monopole-instanton in the path integral comes with three translation zero modes and a Boltzmann suppression factor $e^{-S_0} \sim (NL\Lambda)^{b_0}\ll 1$, where $S_0\approx\frac{8\pi^2}{Ng^2(m_W)}$ is the one-monopole action. This means the typical monopole-instanton separation $d\sim e^{S_0/3}$ is much greater than the monopole diameter $\sim m_W^{-1}$, allowing us to ignore the contribution from paths with overlapping monopole-instanton cores. It also means that we can ignore the effects of $A_4$ exchange.
The proliferation of magnetic charges in the vacuum gives rise to a potential for the photon. This potential which is most conveniently described in terms of the dual photon $\sigma^a$, defined as
\begin{equation}
    \frac{1}{2} \varepsilon_{\mu\nu \rho} \kappa_{ab} F^a_{\mu\nu} = \frac{1}{16\pi}  \partial_\rho \sigma^b .
\end{equation}
Written in terms of $\sigma^a$, the IR behaviour of dYM theory is described to first order in the semiclassical expansion, by the 3d Lagrangian $\mathcal{L}_{3d,\text{dual}}$:
\begin{equation}
    \mathcal{L}_{3d,\text{dual}}=   \frac{1}{2(8\pi)^2} \kappa^{-1}_{ab}\partial_\mu \sigma^a \partial_\mu \sigma^b + \zeta  \sum_{k=1}^{N}\Big [ 1-\cos(\sigma^{k+1}-\sigma^k) \Big ] ,
\end{equation}
where $\sigma^{N+1}\equiv \sigma^1$, and $\kappa^{-1}_{ab}$ is the inverse\footnote{Actually, it should be the pseudoinverse since the eigenvalue corresponding to the $\ell = N$ mode diverges, but the difference is immaterial since the $\ell = N$ mode is unphysical.} of $\kappa_{ab}$, and $\zeta$ is the monopole fugacity:
\begin{equation}
  \zeta \equiv A m_W^3 (g^2 N)^{-2} e^{-8\pi^2/Ng^2(m_W)} .
  \label{eq:fugacity}
\end{equation}
$A(\{m_I\},n_f)$ is an $O(1)$ pre-exponential factor.\footnote{
As an aside, let us note that calculating the pre-factor $A(\{m_I\},n_f)$ is a highly non-trivial open calculation, and has only been performed in the SYM case,  first in \cite{davies00}, and later corrected in \cite{continuity,symallgroups}. This is because it involves matrix determinants in a monopole-instanton background, for which we do not even have an exact analytic expression, as mentioned. We make no attempt to calculate $A$ here. 
\label{fn:fugacity}
}
In the Fourier basis, the 3d dual photon Lagrangian is, to quadratic accuracy in the fields,
\begin{equation}
    \mathcal{L}_{3d,\text{dual}} = \sum_{\ell=1}^{N-1} \Bigg [ \frac{\kappa_{\ell}^{-1} }{(8\pi)^2} |\partial_\mu \tilde{\sigma}^\ell|^2 + \zeta \sin^2 \Big ( \pi\frac{\ell}{N} \Big )| \tilde{\sigma}^\ell|^2  \Bigg ] + O(\tilde{\sigma}^4) ,
    \label{eq:l3d}
\end{equation}
where $\tilde{\sigma}^\ell$ is the discrete Fourier transform of $\sigma^a$:
\begin{equation}
    \tilde{\sigma}^\ell \equiv \frac{1}{\sqrt{N}}\sum_{a=1}^{N}  \omega^{-\ell a} \sigma^a .
\end{equation}
From this expression we can read off the dual photon masses-squared:
\begin{equation}
    m^2_{\sigma,\,\ell} \sim \zeta \sin^2 \Big ( \pi\frac{\ell}{N} \Big ) \, \kappa_{\ell} .
\end{equation}
Let us take $\Lambda N L$ to be sufficiently small so that $\frac{2}{3}n_f\lambda_\ell W_\ell \ll 1 $ can be treated as a small correction for all $\ell$. In that case, we can write a mass-corrected expression for the scaling behaviour of the $k$-wall thicknesses. Recalling \eqref{eq:latticecoupling},
\begin{equation}
    \frac{m_{\sigma,k}}{m_{\sigma,1}} \approx \frac{\sin \pi \frac{ k}{N}}{\sin\frac{\pi}{N}}\left ( \frac{ \lambda_1 }{\lambda_k} \right )^{1/2} \left [
    1+ \frac{n_f}{3} \left ( \lambda_k W_k  - \lambda_1 W_1 \right )
    \right ] .
    \label{eq:kwalls}
\end{equation}
The multiplicative sine factor is the expected tree-level scaling behaviour; the factor of $(\lambda_1/\lambda_k)^{1/2}$ is due to the one-loop corrections in the massless limit, and the factor in the square brackets gives the massive correction. The dependence of $m^2_{\sigma,\ell}$ on $\ell$ in units of $m^2_{\sigma,N/2}$ is graphically depicted in Fig. \ref{fig:mell}.
%%delete the m_\lambda =0 curve.
\subsubsection{Emergent dimension at large $N$: a 4d interpretation of the mass correction}
\label{sec:latticedim}
Let us now consider the large-$N$ limit. To do this, we simultaneously take $N\to \infty$ and $L\to 0$ whilst keeping $NL$ constant so as to stay inside the weak-coupling regime.\footnote{The 't Hooft coupling $g^2N$ is under control in this limit, as can be seen from Equation \eqref{eq:lambdadef}.} This is known as the "Abelian large-$N$ limit" \cite{poppitzreview}. In this setup, we can treat $\frac{\ell}{N}\in [0, 1]$ as though it were on a continuum, and the potential in Equation \eqref{eq:l3d} has an interpretation as the kinetic energy on a latticised and compact fourth dimension, with a \textit{quadratic} (as opposed to \textit{quartic}, as is the case in SYM,) dependence on a lattice momentum $p_y$. But what is the scale of this momentum? Since the mass gap for the dual photon $m^2_{\sigma,\,1}$ vanishes in the large-$N$ limit, the only remaining mass scale to characterise the low-energy theory is $m^2_{\sigma,\,N/2}\equiv m_{N/2}$, the (Debye) mass of the heaviest dual photon:
\begin{subequations}
    \begin{equation}
        \begin{split}
            m_{N/2} \sim m_W \lambda^{-3/2}e^{-1/2\lambda} ,
        \end{split}
    \end{equation}
where
    \begin{equation}
        \begin{split}
            \lambda &\equiv \frac{N g^2 (m_W)}{8\pi^2}\\
            &\approx \lambda_{N/2},
        \end{split}
    \end{equation}
up to small corrections. This allows us to define $p_{y,\ell}$ as an honest-to-goodness lattice momentum:
%%p_y should be p_{y,\ell}
\end{subequations}
\begin{equation}
    p_{y,\ell} \equiv m_{N/2}\sin \Big ( \pi\frac{\ell}{N} \Big ).
\end{equation}
We can also read off the two-point function directly from \eqref{eq:l3d}: Defining $x^M\equiv (\vec{x},y)$, $p_M\equiv(\vec{p},p_y)$, and momentarily disregarding the massive correction,
\begin{equation}
    \int d^4 x e^{i p_M x^M} \langle \sigma(x_M) \sigma(0) \rangle \sim  (  \lambda_\ell \, \vec{p}\, {}^2\!  + \lambda p_y^{2}  )^{-1} .
\end{equation}
We observe that there is a restored Lorentz symmetry which is broken by an anomalous scaling dimension $\Delta = b_0 \lambda$ as $\lambda_\ell \sim p_y^{b_0 \lambda}$.
\par
Put another way, the dual photon coupling $\lambda_\ell$ exhibits logarithmic running in the lattice momentum $p^y\sim \sin ( \pi\frac{\ell}{N} )$:
\begin{equation}
    p_y \frac{d}{ d p_y} \Bigg ( \frac{1}{\lambda_\ell} \Bigg)= p_y \frac{d}{ d p_y} \Big (  b_0 \log \frac{1}{p_y} + (\text{const.}) \Big ) = -b_0 .
\end{equation}
In particular, the scaling behaviour is opposite to that of the $\mathbb{R}^4$ theory (cf. Equation \eqref{eq:lambdadef}):
\begin{equation}
    \mu \frac{d}{ d \mu} \Bigg ( \frac{1}{\lambda(\mu)} \Bigg)= + b_0 .
\end{equation}
\par
We can also show how this analogy can be extended to encompass the mass-correction terms $\sim W_\ell$: The one-loop correction to the coupling due to a single adjoint fermion with mass $m$ in an $SU(N)$ theory on $\mathbb{R}^4$, renormalised at some scale $\mu$ in the $\overline{MS}$ scheme is, (cf. Equation \eqref{regdiagrams}):
\begin{equation}
    \begin{split}
        \mathcal{M}_{\mathbb{R}^4}^{\text{fermion}}(P)  &= -2\int^1_0 \! dx \, x(1-x) \log \Big (\frac{P^2x(1-x)+m^2}{\mu^2}\Big ) \\
        &\sim \begin{cases}
        -\frac{2}{3}\log (\frac{P}{\mu} ) & P^2\gg m^2 ,\\
        -\frac{2}{3}\log( \frac{m}{\mu}  )&P^2\ll m^2 ,
        \end{cases}
    \end{split}
\end{equation}
where $x$ is a Feynman parameter. We can compare this with our result of the contribution in the $\mathbb{R}^3\times S^1$ theory, which can be read off from \eqref{eq:kappaell}:
\begin{equation}
    \begin{split}
        \mathcal{M}_{\mathbb{R}^3\times S^1,\,\ell }^\text{fermion} &= \frac{2}{3}\Big [ \log \Big (\sin \pi \frac{\ell}{N} \Big ) + W_{\ell}\Big ( \frac{m}{m_W}\Big ) \Big ]\\
        &\sim \begin{cases}
        +\frac{2}{3}\log \sin ( \pi \frac{\ell}{N}) & m_W \gg m, \\
         -\frac{2}{3} \log ( \frac{m}{m_W}  ) & m_W \ll m .
        \end{cases}
    \end{split}
\end{equation}
This result is consistent with our interpretation of $p_y$ as a momentum, with the mass correction behaving as we should expect in the $\mathbb{R}^4$ theory, albeit with opposite momentum-scaling behaviour in $p_y$.
\section{Perturbative analysis: theory and practice}
\label{sec:maincalculation}
The remainder of this paper mainly focuses on deriving and calculating loop integrals and Matsubara sums. Our approach is extremely straightforward --- essentially identical to the analysis of a thermal gauge theory at temperatures $T=1/L$, but for the fact that our $S^1$ is spacelike rather than timelike. This means, in particular for the fermions, that the $S^1$ momenta $\omega_n$ assume integer values $\omega_n = \frac{2\pi n}{L}$, rather than half-integer $\omega_n = \frac{2\pi}{L}(n+\frac{1}{2})$. As the calculation is rather involved, our presentation will try to go into as much detail as we can without being overly cumbersome. For the convenience of the reader, we will summarise the contents of Sections \ref{sec:setup} and \ref{sec:wilsonian} at the end of their respective Sections.
\par
Let us start by defining our notation. 
We will use $M,N \in \{1,2,3,4 \}$ for Euclidean indices on $\mathbb{R}^3 \times S^1$, (with $x^4$ the coordinate on $S^1$) and $\mu,\nu \in \{ 1,2,3 \}$ for indices on the $\mathbb{R}^3$. We will use $a,b,c,i,j,k \in \{1, \dots N \}$ to denote Lie algebra indices. 
\par
We also define the (over-complete) Cartan-Weyl basis on $su(N)$:
\begin{subequations}
    \begin{equation}
            (H_i)_{ab} = \delta_{ia}\delta_{ib} = \text{diag} ( 0,\,... \, , \,\overbrace{1}^{i\text{'th}}\,,\, ... \,,\,0 ) \qquad 1\leq i \leq N ,
    \end{equation}
which span the Cartan subalgebra $\mathfrak{t}$. These are accompanied by the raising and lowering operators spanning $\mathfrak{t}^\perp$, the orthogonal complement of $\mathfrak{t}$,
    \begin{equation}
     (E_{\beta_{ij}})_{ab} = \delta_{ai}\delta_{bj} ,
    \end{equation}
for $N$-component vectors $\beta_{ij}$ in the root lattice of $su(N)$, which in our basis are written
	\begin{equation}
		\begin{split}
		    \beta^a_{ij} & \equiv \delta^a_i - \delta^a_j \\ 
	        &= (0, ... \overbrace{1}^{i\text{'th}} , ..., \overbrace{-1}^{j\text{'th}},  ..., 0 ) \,,\qquad 1\leq i\leq j \leq N .
		\end{split}
		\label{eq:rootvecdef}
	\end{equation}
Perhaps a bit idiosyncratically, we say that the subscripts on $\beta_{ij}$ are a set of antisymmetric indices labelling the roots of $su(N)$: $\beta_{ij} = -\beta_{ji}$, and the superscript $a$ denotes its $a$'th vector component. 
\par
$E_{\beta_{ij}}, E_{-\beta_{ij}}$ are respectively raising and lowering operators for the $su(2)$ subalgebra associated with the root $\beta_{ij}$:
\end{subequations}
\begin{subequations}
	\begin{align}
		[ H_i, \, H_j ] = 0\,, \quad
		[ H_k, \,  E_{\beta_{ij}} ]= \beta^k_{ij}  E_{\beta_{ij}}\,, \quad
		[E_{\beta_{ij}}, \, E_{-\beta_{ij}} ] = \sum_k \beta_{ij}^k  H_k ,
	\end{align}
(no sums over $i,j$). We also have
	\begin{align}
	    E_{\beta_{ij}}^\dagger = E_{-\beta_{ij}} = E_{\beta_{ji}} \,,\quad
	    H_k^\dagger = H_k .
	\end{align}
These generators are normalised as:
	\begin{align}
		\trace [H_i H_j] = \delta_{ij} \,,\quad
		\trace[E_\beta E_{-\beta'}] = \delta_{\beta \beta'} .
		\label{matrixnorm}
	\end{align}
	\label{cartan_basis}
\end{subequations}
In the interest of brevity, we will frequently abuse notation and treat $\beta$ as though it were the index on the root space and omit the subscripts $ij$, as we have just done above. To avoid confusion, there will be no implicit sum over $su(N)$ indices unless otherwise specified,.
\par
As a matter of convenience, we normalise the components of $su(N)$-valued fields $\psi$ as:
\begin{subequations}
    \begin{equation}
    	\psi(x^\mu,x^4) = \frac{1}{2} \sum_k \psi^k(x^\mu,x^4) H_k+ \frac{1}{\sqrt{2}}\sum_\beta \psi^\beta(x^\mu,x^4) E_\beta ,
    	\label{csaexpand}
    \end{equation}
obeying hermiticity conditions:
    \begin{equation}
        (\psi^k)^* = \psi^k \, , \qquad
        (\psi^\beta)^* = \psi^{-\beta} ,
    \end{equation}
and constrained by a trace-free condition:
    \begin{equation}
        \sum_k \psi^k (x^\mu,x^4) =0 ,
    \label{tracefree}
    \end{equation}
\end{subequations}
so that the expansion \eqref{csaexpand} is unique although it is written in terms of an over-complete basis. 
\subsection{Formal setup: beginnings}
\label{sec:setup}
Let us start with a 4-dimensional Euclidean $SU(N)$ gauge theory with non-Abelian field strength $\mathcal{F}_{MN}$ and $n_f$ two-component massive adjoint fermions $\lambda_I$. As we are performing a perturbative calculation, the vacuum angle is "invisible" to us, so we might as well set the fermion masses to be real and the topological angle $\theta=0$:
\begin{equation}
    \begin{split}
        \mathcal{L}_{4d} &= \trace \Big [ \frac{1}{2g^2} (\mathcal{F}_{MN})^2 \\
    & + 2i \sum_{I=1}^{n_f} \Big ( \bar{\lambda}_{I}{}_{ \dot \alpha} \bar{\sigma}^{M \dot \alpha \alpha} (\nabla_M  \lambda_{I})_{ \alpha}
    + \frac{m_I}{2} \lambda_{I}{}_{\alpha} \lambda_{I}{}_{\beta} \varepsilon^{\alpha \beta}
    + \frac{m_I}{2} \bar{\lambda}_{I}{}^{\dot \alpha} \bar{\lambda}_{I}{}^{\dot \beta} \varepsilon_{\dot \alpha \dot \beta} \Big ) \Big ] .
    \end{split}
\end{equation}
$\nabla_M$ is the covariant derivative on adjoint-representation fields:
\begin{equation}
    \nabla_M \equiv \dd_M + i[\mathcal{A}_M, \, \cdot \,\,] ,
\end{equation}
and $\bar{\sigma}^M = (\, i\vec{\sigma} \, , \, 1_{2} \, )$ are the Euclidean sigma matrices. 
\par 
Formally integrating out the high-energy (\,$\gtrsim m_W$) degrees of freedom around the centre-symmetric $\Omega$ gives us the effective 3d Lagrangian, \eqref{eq:eftaction}.
\par
We want to explicitly integrate out the high-energy (\,$\gtrsim m_W$) degrees of freedom to obtain the effective 3d Lagrangian, \eqref{eq:eftaction} to find the one-loop corrections to $\kappa_{ab}$, the photon coupling matrix. The methods we use can also be applied almost verbatim to find $\rho_{ab}$, the corrected scalar couplings, as well as $M_{ab}$ the scalar masses. Since these are not as interesting to us, we simply quote their Fourier-transformed results in Equations\eqref{eq:Mell} and \eqref{eq:rhoell}. 
\par
Following Abbott's approach, (e.g., Ref. \cite{abbott},) we use an adapted background field gauge method to calculate vacuum polarisation. This is fairly standard textbook material, but to review: first, we treat the $su(N)$-valued gauge field $A_M$ as the sum of a "classical" background field and a "quantum" high-frequency field:
\begin{equation}
	\mathcal{A}_M = \underbrace{A_M}_{\text{classical}} +  \underbrace{g\,a_M}_{\text{quantum}} .
	\label{Aplusa}
\end{equation}
The normalisation is for convenience. We say that these fields have two complementary expressions of gauge symmetry: for $U,\, \tilde{U} : \mathbb{R}^3\times S^1\to SU(N)$:
\begin{subequations}
	\begin{equation}
		A_M \rightarrow U(A_M +i \dd_M)U^{-1}\,,\quad
		a_M \rightarrow Ua_M U^{-1} \quad \text{(gauge transformation under $U$) },
	\end{equation}
	\begin{equation}
	    a_M \rightarrow \tilde{U}(a_M +i \dd_M)\tilde{U}^{-1}\,, \quad
		A_M \rightarrow \tilde{U}A_M \tilde{U}^{-1}  \quad \text{(gauge transformation under $\tilde{U}$) }.
	\end{equation}
\end{subequations}
Anticipating a 3d and Abelian theory, we take $A_M$ to be Abelian and trivial over $x^4$, and call its field strength $F_{MN}$:
\begin{subequations}
	\begin{equation}
		\dd_4 A_M = 0\,, \qquad [A_M , A_N] = 0 ,
	\end{equation}
	\begin{equation}
		F_{MN} \equiv \dd_M A_N  - \dd_N A_M .
	\end{equation}
\end{subequations}
We want to fix the gauge under $\tilde{U}$ in order to integrate out $a_M$, which, as we will see, are basically $W$-bosons.  To do this, we would impose the condition 
\begin{subequations}
    \begin{equation}
    D^M a_M + ig [a^M,a_M] = 0 ,
\end{equation}
where $D_M$ is the "covariant derivative with connection $A_M$" :
\begin{equation}
    D_M \equiv \dd_M + i [A_M,\, \cdot \,\,] ,
\end{equation}
\end{subequations}
which can be done by adding a Gaussian term to the Lagrangian,
\begin{equation}
\Delta \mathcal{L}_a = \trace \, ( \nabla^M a_M)^2 ,
\end{equation}
and a Lagrangian $\mathcal{L}_c$ for scalar-yet-Grassmannian $su(N)$-valued ghost fields $c$, $\bar{c}$.
\par
Since $A_4$ has a non-zero vev, we must write
\begin{subequations}
    \begin{equation}
        A_4\equiv \frac{\phi}{L} + A_4^0 ,
    \end{equation}
    where $\phi$ is the (constant in $x^\mu$) vev,
    \begin{equation}
        \phi \equiv -iL \log \Omega  ,
    \end{equation}
\end{subequations}
and $A_4^0$ represents the fluctuations around the vev. But we can mostly ignore $A_4^0$ as it is $U(1)^{N-1}$-neutral and therefore not involved in the corrections to $\kappa_{ab}$ at one-loop. Gauge-fixing the centre-symmetric $\Omega$ as in Equation \eqref{eq:omegavev}, $\phi$ has vector components:
\begin{equation}
    \begin{split}
         \phi^k &= \frac{\pi}{N} \sum_{\beta>0} \beta^k  \\
         &= \frac{2\pi}{N} \left (\frac{N+1}{2}-k\right ) ,
    \end{split}
    \label{eq:phi}
\end{equation}
where the sum in Equation \eqref{eq:phi} is over the positive roots. In particular, this means
\begin{equation}
    \phi\vdot \beta_{ij} = \frac{2\pi}{N}(j-i) .
    \label{eq:phidotbeta}
\end{equation}
\par
All together, the gauge-fixed Lagrangian has the form:
\begin{equation}
    \begin{split}
    	\mathcal{L} &= \underset{\textrm{classical fields}}{\underbrace{\mathcal{L}_{cl.}}} + \underset{\textrm{$W$-bosons}}{\underbrace{\mathcal{L}_a+\Delta \mathcal{L}_a}} \\
    	&\qquad+\underset{\textrm{ghosts}}{\underbrace{\mathcal{L}_c}} +  \underset{\textrm{fermions}}{\underbrace{\mathcal{L}_{\lambda}}} + O(\hbar^3) ,
    \end{split}
    \label{eq:Lagrangian}
\end{equation}
where $\mathcal{L}_a$ contains the $\sim Aaa$, $AAaa$ terms upon expanding $\mathcal{L}_{4d}$ in terms of $A_M$ and $a_M$, and similarly for $\mathcal{L}_c$ and $\mathcal{L}_\lambda$. We observe that by choosing $A_M$ to be Abelian, the Abelian parts of each of the quantum fields $a,\lambda,c,\bar{c}$ cannot contribute to $\kappa_{ab}$ at one-loop order, so we may forget about them altogether for the rest of this analysis.
\par
Let us take any $su(N)$-valued field $\psi$ and simultaneously expand in the KK modes and the Cartan-Weyl basis, recalling our convention as in Equation \eqref{csaexpand},
\begin{align*}
	\psi(x^\mu,x^4) &=\frac{1}{2} \sum_{k,z}e^{i\frac{2\pi z}{L} x^4}  \psi^{k,z}(x^\mu) H_k + \frac{1}{\sqrt{2}}\sum_{\beta,z} e^{i\frac{2\pi z}{L} x^4}  \psi^{\beta,k}(x^\mu) E_\beta ,
\end{align*}
so that
\begin{equation}
	iD_4 \psi = \frac{1}{\sqrt{2}} \sum_{z,\beta}  e^{i\frac{2\pi z}{L} x^4} \Big ( \frac{2\pi z + \phi \cdot \beta}{L} \Big ) \psi^{\beta} E_\beta 
	+ \text{(Abelian and $O(\hbar^2)$ parts)},
\end{equation}
so asymptotically, the derivative operator $iD_4$ diagonalises with eigenvalues
\begin{equation}
    iD_4 \to \frac{2\pi z + \phi\vdot \beta_{ij}}{L} \, ,\quad z \,\,\text{integer},
\end{equation}
so that fields in $\mathfrak{t}^\perp$ with charge $\beta$ and circle momentum $2\pi z/L$ of a field with mass $m$ obtains an effective 3d (mass)$^2$:
\begin{equation}
    m^2 +m_W^2 \Big [ N z + (i-j)  \Big ]^2 \geq m_W^2 .
    \label{effmass}
\end{equation}
Thus only $U(1)^{N-1}$-neutral and $x^4$-trivial fields survive in the IR theory at scales $\ll m_W$, consistent with our hypotheses on $A_M$.
\subsubsection*{Summary of Section \ref{sec:setup}:}
We outlined the background field approach to perturbation theory: with an eye toward the infrared theory, we set the background $A_M$ to be $x^4$-trivial and Abelian, and showed that only "quantum fields" proportional to the broken gauge generators may contribute to the corrections of $\kappa_{ab}$ at one-loop order. We further showed that this assumption is self-consistent, because all fields with non-vanishing $x^4$ momentum or carrying charge under the $U(1)^{N-1}$ acquire an effective mass $\geq m_W$ through the Higgs mechanism. 
\subsection{The 1-loop Wilsonian action}
\label{sec:wilsonian}
\par
We can write an expression for the Wilsonian effective action $\Gamma[A]$ by formally integrating out the quantum fields under the path integral sign. Setting the vacuum energy to zero,
\begin{equation}
	\begin{split}
    	\Gamma[A] &= - \log\Big [  \int Da D\bar{c} D c \prod_I^{n_f}(D\bar{\lambda}_I D\lambda_I) \, e^{ - \int\mathcal{L}[A,c,\lambda_I,a] } \Big ] \\
        &= \int_{\mathbb{R}^3}  \Big [ \frac{L}{4g^2} (F_{\mu\nu}^k)^2 + \frac{L}{2g^2} (\dd_\mu A_4^k)^2 \Big ] \\
        &\quad + \sum_{s=0,\frac{1}{2},1} \sum_{f_s} \chi(s) \ftrace \log (-D_{(s)}^2+m_{f_s,s}^2) \\
        & \quad+ \text{(higher loop contributions)} .
	\end{split}
	\label{eq:determinants} 
\end{equation}
There is a lot of notation to define in Equation \eqref{eq:determinants}, but it will make life easier by formatting the problem so that the entire non-trivial part of the calculation is contained in the single expression "$\ftrace \log (-D_{(s)}^2+m_{f_s,s}^2) $," which we will only have to evaluate once to cover all the relevant cases, rather than having to work with massive or massless, spinor, scalar, and vector integrals separately.
\par
"$\ftrace$" refers to the trace over the respective Hilbert spaces, and $-D^2_{(s)}$ is a differential operator defined in Equation (\ref{D2}). The terms on the third row of Equation (\ref{eq:determinants}) are due to the $W$-bosons $a$, ($s=1$,) the gauge ghosts $c,\,\bar{c}$, ($s=0$,) and the fermions $\lambda_I$ ($s=1/2$). The $s=1/2$ term is obtained by doubling then halving the trace-log of the massive Weyl operator:
\begin{equation}
    \begin{split}
        	\sum_{I=1}^{n_f} \frac{1}{2} \ftrace \log (i\bar{\sigma} \! \cdot \! D + im_I)
        	\equiv  \sum_{I=1}^{n_f} \frac{1}{4}  \ftrace\log \Big (-D_{(1/2)}^2 + m_I^2 \Big ) .
        	\label{eq:dhalf}
    \end{split}
\end{equation}
$\sum_{f_s}$ is a sum over flavour indices $I$ when $s=1/2 $; $m_{f_s,s}^2 = 0$ for $s\neq 1/2$. $\chi (s)$ is a pre-factor determined by the statistics of each field:
\begin{equation}
	\begin{split}
		\chi(s) \equiv
		\begin{cases}
    		-1 & (s=0) ,\\
    		-1/4 & (s = 1/2) ,\\
    		+1/2 & (s =1) . \\
		\end{cases}
	\end{split}
    \label{x_def}
\end{equation} 
\par
To define $-D_{(s)}^2$, let $A, B$ denote indices in the spin-$s$ irrep of the (Euclidean) Lorentz group, then
\begin{equation}
	\begin{split}
		(-D_s^2)_{AB} &= -D_M D^M \delta_{AB} +  F^k_{MN}(H^{\text{adj.}}_k) (\sigma^{(s)}_{MN})_{AB} \\
		&= \Big [ (i\dd_\mu)^2 + \Big( i\dd_3+ \frac{\phi^k }{L}(H_k^{\text{adj.}})  \Big )^2\,  \Big ] (\delta^{(s)})_{AB}\\
		&\quad +(\Sigma_{1}^{(s)})_{AB}+ (\Sigma_{2}^{(s)})_{AB}+ (\Sigma_{F}^{(s)})_{AB}  ,
	\end{split}
    \label{D2}
\end{equation} 
(with an implicit sum over $k$,) where $(H_k^{\text{adj.}}) \equiv [H_k,\,\cdot\,\,]$, and
\begin{subequations}
    \begin{align}
    	(\Sigma_{F}^{(s)})_{AB} &\equiv   F^k_{MN}(H_k^{\text{adj.}}) (\sigma^{(s)}_{MN})_{AB} , \label{eq:vertexf} \\
    	(\Sigma_{1}^{(s)})_{AB}  &\equiv  - i A^k_M \overleftrightarrow{\partial}^M (H_k^{\text{adj.}}) (\delta^{(s)})_{AB} , \label{eq:vertex1} \\
    	(\Sigma_{2}^{(s)})_{AB}  &\equiv[ A^k_M (H_k^{\text{adj.}})]^2 (\delta^{(s)})_{AB} , \label{eq:vertex2} 
    \end{align}
\end{subequations} (again, with an implicit sum over $k$). $\Sigma_{1}^{(s)}$ and $\Sigma_{2}^{(s)}$ are respectively the 3- and 4- point interactions of a charged adjoint field, and $\Sigma_F^{(s)}$ is the spin-field coupling term responsible for asymptotic freedom in non-Abelian theories.  $\delta^{(s)}$ and $\sigma_{MN }^{(s)}$ are respectively the identity matrix and the generators of rotations in the spin-$s$ representation. Explicitly, (and abusing notation slightly by mixing indices,)
\begin{equation}
		(\sigma_{MN}^{(s)})_{AB}=
		\begin{cases}
		    \, 0 & (s=0) , \\
    		\,\frac{i}{4} ( \bar{\sigma}_{[M} \sigma_{N]} )_{AB}  & (s=1/2), \\
    	    \,-i(\delta_{A M}\delta_{NB}\! \!-\! \!\delta_{A N }\delta_{ MB}) \quad & (s=1)  .
    	\end{cases}
\end{equation}
so the one-loop correction to the Wilsonian can be written, to quadratic order in $A_M$:
\begin{equation}
    \begin{split}
	    \ftrace \log \Big ( \frac{-D^2_{s} +m_s^2}{-\dd^2 + m_s^2} \Big ) =  &\ftrace \Big ( \frac{ \Sigma^{(s)}_2}{-\dd^2+ m_s^2 } \Big ) \\  - \frac{1}{2} & \ftrace \Big [ \Big ( \frac{\Sigma^{(s)}_1}{-\dd^2+ m_s^2}\Big )^2 + \Big ( \frac{\Sigma^{(s)}_F}{-\dd^2+ m_s^2}\Big )^2 \Big ] + O(A^3) . \label{eq:twopoint}
	\end{split}
\end{equation}
There is no $\sim \Sigma_F \Sigma_1$ cross-term because the trace of $\sigma^{(s)}_{MN}$ vanishes.
Expanding in a Fourier basis to quadratic order in the fields, Equation \eqref{eq:determinants} becomes
\begin{equation}
    \begin{split}
         \Gamma[A_M^k;\mu]  =  2\sum_{a,b}\int \!\frac{d^3p}{(2\pi)^3}  \Big [ &A^{\mu a} \kappa_{ab}  ( p^2 \delta_{\mu\nu} - p_\mu p_\nu) A^{\nu b} \\
          + & A_4^a \Big ( p^2\rho_{ab} + \frac{L}{2g^2} M^2_{ab} \Big ) A_4^b  \Big ] + O(A^3) .
    \end{split}
    \label{eq:pwilsonian}
\end{equation}
We note in passing that the GPY potential $V[\Omega]$ still appears in Equation \eqref{eq:pwilsonian} through $M^2_{ab}$, its second derivative.
\par
Now we are ready to draw some Feynman diagrams. Let $p^M = (p^\mu, 0)$ denote the external momentum of $A_M$, and for convenience, define an  effective loop momentum $K^M_{(\beta,z)}$.
\begin{equation}
    K^M\equiv  K^M_{(\beta,z)} \equiv \Big ( k^\mu, \frac{2\pi z +\phi\cdot \beta}{L} \Big ) .
\end{equation} 
Using Equations \eqref{eq:vertexf}, \eqref{eq:vertex1}, \eqref{eq:vertex2}, and \eqref{eq:twopoint} and reading off from \eqref{eq:pwilsonian}, we can write down the corrections to the $\sim A_M^a A_N^b$ term in the action:
\begingroup
\begin{subequations}
    \begin{align}
%--seagull--%
		(\Pi_2^{(s)})_{MN}^{ab}&\equiv \frac{1}{2} \frac{\delta^2}{\delta A^a_M\delta A^b_N} \ftrace \Big ( \frac{ \Sigma^{(s)}_2}{-\dd^2+ m_s^2 } \Big )
		\nonumber\\
		& = \frac{1}{L}\sum_{z, \beta }   \int \! \frac{d^3 \! k}{(2\pi)^3}
			d(s)  \beta^a \beta^b \delta_{MN} \Bigg [ \frac{1}{(K^2+\Delta_s^2)} +  \frac{(1-2x)^2 p^2}{2(K^2+\Delta_s^2)^2} \Bigg ], \label{seagull}  \\
%%--current loop--%%
		(\Pi_1^{(s)})_{MN}^{ab} &\equiv \frac{1}{2} \frac{\delta^2}{\delta A^a_M\delta A^b_N}  \Bigg [ -\frac{1}{2} \ftrace  \Big ( \frac{ \Sigma^{(s)}_1}{-\dd^2+ m_s^2 } \Big )^2 \Bigg ] 
		 \nonumber\\
		&=-\frac{1}{L}\sum_{z, \beta }   \int \! \frac{d^3 \! k}{(2\pi)^3} \frac{1}{2}  d(s)   \beta^a \beta^b \int^1_0 \! dx   \frac{4 K_M  K_N +(1-2x)^2 p_M  p_N }{(K^2 + \Delta_s)^2} , \label{current} \\
%%--spin-field--%%
		(\Pi_F^{(s)})_{MN}^{ab} &\equiv \frac{1}{2} \frac{\delta^2}{\delta A^a_M\delta A^b_N}  \Bigg [ -\frac{1}{2}\ftrace   \Big ( \frac{ \Sigma^{(s)}_F}{-\dd^2+ m_s^2 } \Big )^2  \Bigg ]
		 \nonumber \\
		&=-\frac{1}{L}\sum_{z, \beta }   \int \! \frac{d^3 \! k}{(2\pi)^3} c(s)   \beta^a \beta^b \int^1_0 \! dx  \frac{ 2(p^2\delta_{MN} -p_M  p_N   )}{(K^2 + \Delta_s)^2}. \label{spinfield}
	\end{align}
	\label{eq:MNdiagrams}
\end{subequations}
\endgroup
\begin{figure}[]
  \begin{subfigure}{0.31\textwidth}
    \includegraphics[width=\linewidth]{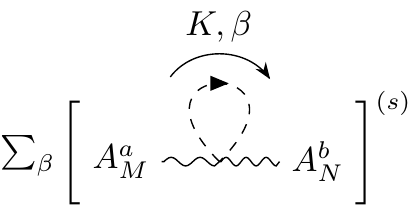}
    \vspace*{0.16cm}
    \caption{($\Pi_2^{(s)})_{MN}^{ab}$} \label{fig:1a}
  \end{subfigure}%
  \hspace*{\fill}   % maximize separation between the subfigures
  \begin{subfigure}{0.31\textwidth}
    \includegraphics[width=\linewidth]{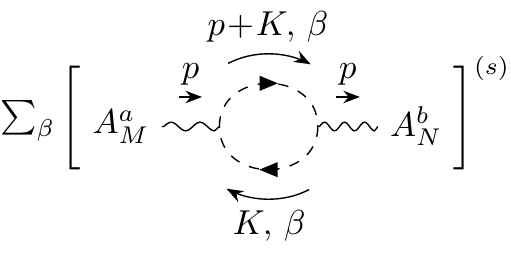}
    \caption{($\Pi_1^{(s)})_{MN}^{ab}$} \label{fig:2b}
  \end{subfigure}%
  \hspace*{\fill}   % maximizeseparation between the subfigures
  \begin{subfigure}{0.31\textwidth}
    \includegraphics[width=\linewidth]{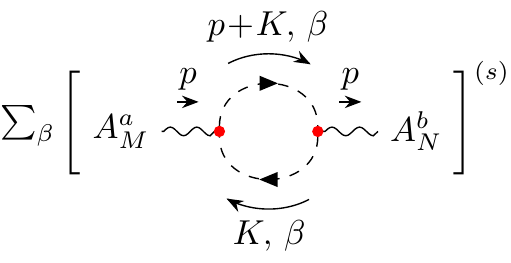}
    \caption{($\Pi_F^{(s)})_{MN}^{ab}$} \label{fig:3c}
  \end{subfigure}
\caption{Representations of the loop integrals in Equation \eqref{eq:MNdiagrams} in terms of Feynman diagrams. The $\Sigma_F$ vertex is distinguished from the $\Sigma_1$ vertex with a (red) dot.} \label{fig:1}
\end{figure}
These integrals are pictorially represented by the Feynman diagrams in Fig. \ref{fig:1}. In each of the integrals above we have employed a Feynman parameter $x$, and shifted our loop momentum $K^M \to K^M -x p^M$. We have also defined an effective (mass)$^2$, $\Delta_s$ (not to be confused with the 3d effective (mass)$^2$ in Equation \eqref{effmass},)
\begin{equation}
	\Delta_s \equiv m_s^2 +p^2 x(1-x) ,
\end{equation}
We have further defined $d(s)$, the number of spin states in the spin-$s$ representation, and $c(s)$, the spin-field coupling coefficient:\footnote{Note that $d(1/2) = 4$ for us, because we doubled the number of polarisations in Equation \eqref{eq:dhalf}; this is already compensated for by an additional factor of $1/2$ in front of the fermion determinant in Equation \eqref{eq:determinants}.}
\begin{equation}
    	d(s) \equiv\trace (\delta^{(s)}) = 
    	\begin{cases}
    		1 & (s=0) , \\
    		4 & (s = 1/2) ,\\
    		4 & (s =1) , \\
    	\end{cases}
     \,  \qquad c(s) \equiv\trace (\sigma^{(s)}_{MN}\sigma^{(s)}{}^{MN}) = 
    	\begin{cases}
    		0 & (s=0), \\
    		1 & (s = 1/2) ,\\
    		2 & (s =1), \\
    	\end{cases}
    	\label{d_def}
\end{equation}
where the traces are over the spin indices, which we have omitted. The rest of our report will be largely concerned with evaluating these three integrals.
\subsubsection*{Summary of Section \ref{sec:wilsonian}:}
We introduced some formal notation to write down the one-loop effective action in a more compact form, Equation \eqref{eq:determinants}. This allowed us to write the integrals of each of $a,\lambda, c$ in terms of the loop integrals $\Pi^{(s)}_{2}$, (Equation \eqref{seagull},) $\Pi^{(s)}_{1}$, (Equation \eqref{current},) and $\Pi^{(s)}_F$ (Equation \eqref{spinfield}). As we will see, the evaluation of these integrals are by no means a trivial task, but we will make them much more tractable with a handful of clever manipulations.
\subsection{Outline of the calculation}
%%%say that integrals are hards
\label{sec:outline}
We have written the integrals in \eqref{eq:MNdiagrams} to superficially respect the Euclidean Lorentz group $SO(4)$. But to evaluate them, we must rewrite \eqref{eq:MNdiagrams} to reflect the broken rotational symmetry $SO(4)\to SO(3)$. Symmetry considerations tell us that averaging $K_M K_N$ must give:
\begin{subequations}
\begin{equation}
	( \overline{K_M K_N})_{(\beta,z)} = \frac{k^2}{3}\delta_{\mu\nu}\mathcal{P}^{\mu\nu}_{MN} + \Big(\frac{2 \pi z+\phi\cdot\beta}{L} \Big )^2\mathcal{P}^{44}_{MN} ,
\end{equation} where $\mathcal{P}^{\mu\nu}_{MN}$ and $\mathcal{P}^{44}_{MN}$ are projection operators to $\mathbb{R}^3$ and $S^1$ respectively:
    \begin{align}
        \mathcal{P}^{\mu\nu}_{MN} &\equiv  \delta_{M}^\mu \delta_{N}^\nu , \label{projectorMN} \\
        \mathcal{P}^{44}_{MN} &\equiv \delta_{M}^4 \delta_{N}^4 ,
        \label{projector44}
    \end{align}
\end{subequations}
Integrating over the angular coordinates and summing the three graphs in Equation \eqref{eq:MNdiagrams}, we get
\begin{subequations}
    \begin{equation}
       \begin{split}
           (\Pi^{(s)})_{\mu\nu}^{ab} &\equiv \sum_{\mathcal{I}=2,1,F} (\Pi_\mathcal{I}^{(s)})_{\mu\nu}^{ab} \\
           &=\sum_{z, \beta } \beta^a \beta^b  \! \int^1_0 \! dx  \int^\infty_0 \frac{d(kL)}{2\pi^2} (kL)^2 \Bigg \{ \Big [ \frac{(1-2x)^2}{2}d(s) -2c(s) \Big ] (p^2\delta_{\mu\nu}- p_\mu p_\nu )S_1(b,\omega_s L ) \\ 
           &\qquad \qquad\qquad \qquad\qquad \qquad \qquad+ \frac{d(s)}{L^2} \delta_{\mu \nu } \Big [ S_0(b,\omega_s L ) -\frac{2}{3} (kL)^2 S_1(b,\omega_s L ) \Big ]\Bigg \} .
       \end{split}
       \label{eq:munudiagrams}
    \end{equation}
We also write out the $(44)$ part, which are needed to renormalise:
    \begin{equation}
        \begin{split}
            (\Pi^{(s)})_{44}^{ab} &\equiv \sum_\mathcal{I} (\Pi_\mathcal{I}^{(s)})_{44}^{ab} \\
            &=\sum_{z, \beta } \beta^a \beta^b  \! \int^1_0 \! dx  \int^\infty_0 \frac{d(kL)}{2\pi^2} (kL)^2 \Bigg \{ \Big [ \frac{(1-2x)^2}{2}d(s) -2c(s) \Big ] p^2 \, S_1(b,\omega_s L ) \\ 
           &\qquad \qquad\qquad \qquad\qquad \qquad \qquad \qquad \qquad + \frac{d(s)}{L^2} \Big [ S_0(b,\omega_s L ) - 2 S_2(b,\omega_s L ) \Big ]\Bigg \} .
        \end{split}
        \label{eq:44diagrams}
    \end{equation}
where $(\Pi^{(s)})_{\mu \nu }^{ab}$ and $(\Pi^{(s)})_{44}^{ab}$ are defined in the obvious way:
    \begin{equation}
        (\Pi^{(s)})_{MN}^{ab} \equiv (\Pi^{(s)})_{\mu \nu }^{ab}  \mathcal{P}^{\mu \nu}_{MN} + (\Pi^{(s)})_{44}^{ab}  \mathcal{P}^{44}_{MN} ,
    \end{equation}
and we have also defined:
    \begin{equation}
        \omega_s \equiv \sqrt{k^2+\Delta_s} \, , \qquad 
       b \equiv \phi\cdot \beta ,
    \end{equation}
and dimensionless sums over the KK modes, $S_{0,1,2}$:  
    \begin{equation}
          	S_1(b,\omega L) \equiv\sum_{n\in\mathbb{Z}} \frac{1}{[(2\pi n+b)^2 + (\omega L)^2]^2} \, , \quad  S_2(b,\omega L) \equiv \sum_{n\in\mathbb{Z}} \frac{(2\pi n + b)^2}{[(2\pi n+b)^2 + (\omega L)^2]^2} .
    \end{equation}
The third sum, $S_0$, is a standard result. It can be evaluated exactly by e.g., Matsubara summation:
    \begin{equation}
        \begin{split}
        	S_0(b,\omega L) &\equiv\sum_{n\in\mathbb{Z}} \frac{1}{(2\pi n+b)^2 + (\omega L)^2} \\
        	&= \frac{1}{2 \omega L} + \frac{1}{2 \omega L} Re \Big ( \frac{1}{e^{L\omega+ib}-1} \Big )\\
        	&\equiv I_0^{\text{vac.}}(\omega L) + \delta I_0(b,\omega L) .
        	\label{matsubara}
    	\end{split}
    \end{equation}
\end{subequations}
Where we have defined a function $I^{\text{vac.}}_0\equiv  \frac{1}{2 \omega L} $ that falls off as a negative power in $\omega L$, and another, $\delta I_0 \equiv  \frac{1}{2 \omega L} Re ( \frac{1}{e^{L\omega+ib}-1} )$, that falls off exponentially.
\par
Since the summand of $S_0$ is monotone decreasing in $|n|$, differentiation commutes with summation, so $S_{1,2}$ can be trivially evaluated by taking derivatives of both sides of Equation \eqref{matsubara}:
\begin{subequations}
    \label{eq:sums}
    	\begin{align}
    		S_1(b,\omega L) &= -\frac{\dd}{\dd{(\omega L)^2}} S_0(b,\omega L)\nonumber  \\ 
    		&\equiv  I^{\text{vac.}}_1(\omega L) + \delta I_1(b,\omega L),  \label{s_sums1} \\
    		\nonumber\\
    		S_2(b,\omega L) &=  \frac{\dd}{\dd{(\omega L)^2}} \Big [ (\omega L)^2 S_0(b,\omega L) \Big ] \nonumber \\
    		&\equiv  I^{\text{vac.}}_2(\omega L) + \delta I_2(b,\omega L),  \label{s_sums2} 
    	\end{align}
\end{subequations}
where $ I^{\text{vac.}}_{1,2}$ and $\delta I_{1,2}$ are defined in terms of derivatives of $I_0^{\text{vac.}}$ and $\delta I_0$ respectively, in the obvious ways as suggested by the notation. The point is that we can split the integrals in Equation \eqref{eq:munudiagrams}:
\begin{equation}
    (\Pi^{(s)})_{\mu\nu}^{ab} = (\Pi^{(s), \textrm{vac.}})_{\mu\nu}^{ab} + (\delta \Pi^{(s)})_{\mu\nu}^{ab} ,
    \label{eq:splitting}
\end{equation}
by collecting the $ I^{\text{vac.}}_{0,1,2}$ terms into $ (\Pi^{(s), \textrm{vac.}})_{\mu\nu}^{ab}$, and the $\delta I_{0,1,2}$ terms into $(\delta \Pi^{(s)})_{\mu\nu}^{ab}$, and similarly for $(\Pi^{(s)})_{44}^{ab}$. We will call these the \emph{vacuum integral} and \emph{pseudo-thermal integral} contributions respectively, and we consider them separately in the following.
\par 
The basic idea is this: we can see by the asymptotics that the $(\Pi^{(s),\text{vac.}})_{MN}^{ab}$ integrals remain unchanged in the $L\to \infty$ limit. This means we can evaluate those integrals in terms of the familiar loop integrals in $\mathbb{R}^4$, in a way we show explicitly. Obviously these integrals are UV divergent, but they can be renormalised in the $\overline{MS}$ scheme in the usual way. On the other hand, the $SO(4)$-breaking, $L$-dependent parts of $(\Pi^{(s)})_{MN}^{ab}$ are contained entirely within $(\delta \Pi^{(s)})_{MN}^{ab}$ : the exponential decay of the $\delta I_{0,1,2}$ means that the integrands of $(\delta \Pi^{(s)})_{\mu\nu}^{ab}$ are uniformly convergent in $kL$. Then we may use the identities \eqref{s_sums1}, \eqref{s_sums2} to integrate by parts in $kL$ and obtain a much more tractable expression.
\subsubsection{The vacuum integrals}
\label{sec:vac}
We can evaluate the loop integrals in $\Pi^{(s), \textrm{vac.}}$ by "undoing" an integral over an auxiliary continuous variable $k_4$. For example, (defining $\omega \equiv \sqrt{k^2 +\Delta}$ for positive $\Delta$,)
\begin{equation*}
    \begin{split}
        \int \! \frac{d^3k}{(2\pi)^3} L I^{\text{vac.}}_0 &=   \int \! \frac{d^3k}{(2\pi)^3}\frac{1}{2\omega} \\
         &= \int \! \frac{d^3 k}{(2\pi)^3} \!\! \int^{\infty}_{-\infty} \frac{dk_4}{2\pi} \frac{1}{(k_4)^2 + k^2 + \Delta} ,
     \end{split}
\end{equation*}
thus mapping the integral over $k \in \mathbb{R}^3$ to one over $\tilde{k} \in \mathbb{R}^4$. Then we regulate the expression by taking the analytic continuation to $d\equiv4-\varepsilon$ dimensions. In summary:
\begin{subequations}
    \begin{align}
        	\int \! \frac{dk}{2\pi^2}k^2 \, L I^{\text{vac.}}_1
        	&\longrightarrow \mu^{-\varepsilon} \int \! \frac{d^d  \tilde{k}}{(2\pi)^d} \frac{1}{(\tilde{k}^2 + \Delta)^2} , \\
      	    \int \! \frac{dk}{2\pi^2}k^2\, L^3 I^{\text{vac.}}_2
      	    &\longrightarrow \mu^{-\varepsilon} \int \! \frac{d^d  \tilde{k}}{(2\pi)^d} \frac{(k_4)^2}{(\tilde{k}^2 + \Delta)^2} , \\
            \int \! \frac{dk}{2\pi^2} k^2 L I^{\text{vac.}}_0  &\longrightarrow \mu^{-\varepsilon} \int \! \frac{d^d  \tilde{k}}{(2\pi)^d} \frac{1}{(\tilde{k}^2 + \Delta)} .
    \end{align}
\end{subequations}
The expressions on the LHS are the relevant $\mathbb{R}^3$ integrals, and $\mu$ is the $\overline{MS}$ scale of the theory. "$\longrightarrow$" means "analytically continues to." On the other hand, we also have the following series of relations under the integral sign:
\begin{equation}
    \begin{split}
	    \frac{\tilde{k}^2}{d}\delta^{MN}|_{\mathbb{R}^d} \equiv & \,\overline{(\tilde{k}^M \tilde{k}^N)}\,|_{\mathbb{R}^d} \\
	    \longleftarrow & \,  \, [\,\overline{(k^{\mu}k^{\nu})} \mathcal{P}_{MN}^{\mu\nu} + (k_4)^2\mathcal{P}_{MN}^{44}]\, |_{\mathbb{R}^3\times S^1}\\
	    =&\, \Big [\,\frac{k^2}{3} \mathcal{P}_{MN}^{\mu\nu} \delta_{\mu\nu} + (k_4)^2\mathcal{P}_{MN}^{44} \Big]\, |_{\mathbb{R}^3\times S^1} .
	\end{split}
\end{equation}
where $\mathcal{P}_{MN}^{44}, \mathcal{P}_{MN}^{\mu \nu}$ are the projectors defined in \eqref{projectorMN} and \eqref{projector44}. Combining these expressions, the vacuum integrals can be re-written as integrals in $d=(4-\varepsilon)$ dimensions by restoring the $SO(4)$ symmetry:
\begin{equation}
    \begin{split}
        (\Pi^{(s), \textrm{vac.}})_{MN}^{ab} &\equiv \sum_\mathcal{I} (\Pi_\mathcal{I}^{(s), \textrm{vac.}})_{MN}^{ab} \\
	    &= \sum_{\beta } \beta^a \beta^b  \int^1_0 \! dx 
	    \int \! \frac{d^d  \tilde{k}}{(2\pi)^d}
	    \frac{\mu^{4-d}}{(\tilde{k}^2+\Delta_s)^2} \Bigg \{ \Big [ \frac{d-2}{2}\tilde{k}^2 + \Delta_s \Big ] d(s) \delta_ {MN} \\
	    &\qquad\qquad\qquad + \Big [ \frac{(1-2x)^2}{2}d(s)-2c(s) \Big ] \Big [ ( p^2 \delta_{\mu\nu} - p_\mu p_\nu )\mathcal{P}^{\mu\nu}_{MN} + p^2\,\mathcal{P}^{44}_{MN}\Big ]
	    \Bigg \} .
    \end{split}
\end{equation}
Expanding in powers of $1\gg\varepsilon>0$, it is easy to regulate the $\tilde{k}$ integral to get a convergent result. The (Abelian part of the) UV counterterm $\delta Z_s\trace F_{MN}F^{MN}$ contributes, diagrammatically, 
\begin{subequations}
    \begin{equation}
	\begin{split}
		  \Bigg [\feynmandiagram [horizontal=b to c, layered layout,inline=(b.base)] {
		 	a [particle ={\(A^a_M\)}] -- [boson] b [crossed dot]
			-- [boson] c [particle ={\(A^b_N\)}],
		}; \Bigg ]^{\textrm{CT},(s)} 
			&=  \delta Z_s \sum_{\beta}\beta^a\beta^b \Big [ ( p^2 \delta_{\mu\nu} - p_\mu p_\nu )\mathcal{P}^{\mu\nu}_{MN} + p^2\,\mathcal{P}^{44}_{MN}\Big ] .
	\end{split}
	\label{countertermdiagram}
\end{equation}
So for each $s$, we choose
\begin{align}
	-\delta Z_s&\equiv \frac{1}{32\pi^2} \Big ( \frac{d(s)}{3} -4c(s) \Big)\Big ( \frac{2}{\varepsilon} -\gamma +\log 4\pi\Big ) ,
\end{align}
\end{subequations}
and the sum of the three \textit{regulated} vacuum integrals is therefore:
\begin{equation}
    \begin{split}
         (\tilde{\Pi}^{(s), \textrm{vac.}})_{MN}^{ab} &\equiv  (\Pi^{(s), \textrm{vac.}})_{MN}^{ab} + (\textrm{counterterms}) \\ 
        &= \sum_\beta \frac{\beta^a \beta^b}{32\pi^2}\int^1_0 \! dx  \Big [ d(s) (1-2x)^2 - 4c(s) \Big ] \\
        & \qquad\qquad\qquad\qquad \times
        \Big[ (p^2 \delta_{\mu\nu} - p_\mu  p_\nu) \mathcal{P}^{\mu\nu}_ {MN} +  p^2 \mathcal{P}^{44}_ {MN} \Big]\log \Big ( \frac{\mu^2}{\Delta_s}\Big ).
    \end{split}
	\label{regdiagrams}
\end{equation}
\subsubsection{The pseudo-thermal integrals}
\label{sec:qtherm}
Now we consider the pseudo-thermal integrals. Using Equations \eqref{s_sums1},  \eqref{s_sums2}, we can simplify the loop integrals immensely by integrating by parts by changing variables $\frac{\dd}{\dd (\omega L)^2} = \frac{1}{2kL} \frac{\dd}{\dd (kL)} $. We find that all boundary terms vanish, and the results are, in summary,
\begin{subequations}
	\begin{alignat}{2}
		&\int^\infty_0 \! d(kL)\, (kL)^2 \delta I_1(b,\omega L) 	=&&\frac{1}{2} \int^\infty_0 \! d(kL)\, \delta I_0(b,\omega L) , \\
		&\int^\infty_0 \! d(kL)\, (kL)^4 \delta I_1(b,\omega L) 	=&&\frac{3}{2} \int^\infty_0 \! d(kL)\, (kL)^2 \delta I_0(b,\omega L) ,\\
		&\int^\infty_0 \! d(kL)\, (kL)^2 \delta I_2(b,\omega L)  = -&&\frac{1}{2} \int^\infty_0 \! d(kL)\, (\omega L)^2 \delta I_0(b,\omega L) .
	\end{alignat}
\end{subequations}
Plugging into Equations \eqref{eq:munudiagrams}, \eqref{eq:44diagrams}, the pseudo-thermal integrals may be written: 
\begin{equation}
    (\delta \Pi^{(s)})_{\mu\nu}^{ab}=\sum_{\beta} \frac{\beta^a \beta^b}{8\pi^2}  \int^1_0 \! dx \, \Big [ d(s)(1-2x)^2 
	-4c(s)\Big ](p^2\delta_{\mu \nu} -p_\mu p_\nu)	R^b_0(\sqrt{\Delta_s}L) ,
	\label{resultmunu}
\end{equation}
Where we have defined\footnote{The integral in the second line can be carried out by expanding in series in $|e^{-\sqrt{(kL)^2+\Delta L^2}-ib}| < 1$.}
\begin{equation}
    \begin{split}
        R^b_0(\sqrt{\Delta}L) &\equiv  \int^\infty_0 \! d(kL) \cdot \delta I_0\Big  ( b, \sqrt{(kL)^2+\Delta L^2}\Big  ) \\
        &= \int^\infty_0\frac{d(kL)}{2 \sqrt{(kL)^2+\Delta L^2}} Re \Big ( \frac{1}{e^{\sqrt{(kL)^2+\Delta L^2}+ib}-1} \Big ) \\
    &= \sum_{n=1}^{\infty}K_0 (n\sqrt{\Delta}L) \cos (n \phi \cdot \beta_{ij}) ,
    \label{R0def}
    \end{split}
\end{equation}
where $K_0$ is the modified Bessel function of order $0$. This represents the only remaining non-trivial sum, as far as the corrections to $\kappa_{ab}$ are concerned. We have not given an expression for $(\delta \Pi^{(s)})_{44}^{ab}$ as it is not needed to find $\kappa_{ab}$. 
\par
Summing the result with the vacuum contribution,
    \begin{equation}
    	 \begin{split}
    	     &(\Pi^{(s),\textrm{vac.}})_{\mu\nu}^{ab} + (\delta \Pi^{(s)})_{\mu\nu}^{ab}  + \text{(counterterms)}\\
    	    &=\frac{(p^2\delta_{\mu \nu} -p_\mu p_\nu)}{8\pi^2}  \int^{1}_0 dx \Big [ d(s) (1-2x)^2 -4c(s)\Big ]  \Big (\mathcal{R}_0^{ab}(\sqrt{\Delta_s} L) + \delta^{ab} N \log \frac{\mu}{\sqrt{\Delta_s}}\Big) ,
    	 \end{split}
    \end{equation}
where we have defined
    \begin{equation}
        \begin{split}
              \mathcal{R}_0^{ab}(\sqrt{\Delta} L)  & \equiv  \sum_{i,j} \beta^a_{ij} \beta^b_{ij} R_0^{\phi \cdot \beta_{ij}}(\sqrt{\Delta} L) \\ 
              &=  \sum_{i,j} \beta^a_{ij} \beta^b_{ij}  \sum_{n=1}^{\infty}K_0 (n\sqrt{\Delta}L) \cos (n \phi \cdot \beta_{ij}) .
        \end{split}
        \label{eq:r0matrix}
    \end{equation}
In Appendix \ref{sec:r0} we explicitly show that 
\begin{equation}
    \begin{split}
         R_0^b (\sqrt{\Delta} L) = \frac{1}{2}\log \frac{\sqrt{\Delta} L}{4\pi } + \tilde{R}_0^b(\sqrt{\Delta} L) ,
    \label{eq:logr0}
    \end{split}
\end{equation}
where $\tilde{R}_0^b(t)$ is a pure function that has a power series expansion around $t=0$ for fixed $b\in(0,2\pi)$. We know that Equation \eqref{eq:logr0} must be true because the running of the coupling $g^2$ must freeze out at scales below $m_W$. Equation \eqref{eq:logr0} allows us to disregard the $p^2$-dependence in $R_0^b(\sqrt{\Delta}L)$ as higher-derivative corrections, and integrate over the Feynman parameter $x$ trivially. Recalling Equations \eqref{d_def} and \eqref{x_def},
\begin{subequations}
           \begin{equation}
            \sum_{s=0,1}  \chi(s) \Big (  4 \, c(s)- \frac{d(s)}{3} \Big ) = \frac{11}{3} ,
        \end{equation}
and
        \begin{equation}
            \chi(1/2) \Big (  4 \, c(1/2)- \frac{d(1/2)}{3}  \Big ) =- \frac{2}{3} .
        \end{equation}
\end{subequations}
we have
\begin{equation}
    \kappa_{ab} = \frac{m_W^{-1}}{16\pi} \Big[ \frac{8\pi^2}{N g^2(\frac{4\pi}{L})} \delta_{ab}+ \frac{1}{N}\sum_{i,j} \beta^a_{ij}\beta^b_{ij} \Big (\frac{11}{3}\tilde{R}^{\phi\cdot\beta_{ij} }_0(0) - \frac{2}{3} \sum_{I=1}^{n_f} \tilde{R}^{\phi\cdot\beta_{ij} }_0(m_I L) \Big ) \Big ] .
    \label{eq:kab}
\end{equation}
An expression for $\tilde{R}_0^{\phi\cdot\beta_{ij}}(t)$ is derived in Equation \eqref{eq:r0tilde}.
All that remains now is to diagonalise Equation \eqref{eq:kab}.
\subsection{The sums over $\beta$: linear algebra on the root lattice}
Let us consider the sums over the root vectors $\beta$. It is not hard to show by standard Fourier analysis that, for any integer $n$,
\begin{equation}
    \begin{split}
        C^{ab}_{n} :&= \sum_{i,\,j} \beta^a_{ij} \beta^b_{ij} \cos \Big(\frac{2\pi}{N}n(i-j)\Big ) \\
        &= 2(N\delta_{N\equiv n}\delta^{ab}-1) \cos \Big(\frac{2\pi}{N}n(a-b) \Big ) ,
    \end{split}
    \label{eq:cnab}
\end{equation}
where
\begin{equation}
\delta_{n \equiv k} = 
\begin{cases} 
    1 & n \equiv k , \\ 
    0 & n \not\equiv k . 
\end{cases}
\end{equation}
The relation "$\equiv$" is to be understood here as equality in the mod $N$ sense (we instead use "$:=$" to denote "is defined to be" for this subsection).
\par
The matrix in Equation \eqref{eq:cnab} is diagonalised by the (trace-free) eigenvectors $u_\ell$ with vector components:
\begin{equation}
    (u_\ell)^b := e^{ i \frac{2\pi}{N} \ell b},\qquad 1\leq \ell \leq N-1 ,  
\end{equation}
and have eigenvalues indexed by $\ell$:
\begin{equation}
    \begin{split}
	  \sum_{b=1}^N C^{ab}_n (u_\ell)^b 
	  &= N(2 \delta_{n\equiv N} - \delta_{n\equiv \ell} - \delta_{n\equiv N -\ell}) (u_\ell)^a .
	 \end{split}
	 \label{C_eigenvalue}
\end{equation}
Plugging Equation \eqref{C_eigenvalue} into Equation \eqref{eq:r0matrix}, and recalling $m_W=\frac{2\pi}{NL},$ we can read off the eigenvalues $\mathcal{R}_{0\ell}$ of $\mathcal{R}_0^{ab}$:
\begin{equation}
    \begin{split}
        \mathcal{R}_{0\ell}(mL) &=  N \sum_{p=1}^{\infty}  \Big \{  2K_0 \Big ( 2\pi p \frac{m}{m_W} \Big )  - K_0 \Big [ 2\pi \Big (p- \frac{\ell}{N} \Big ) \frac{m}{m_W} \Big ] \\
        &\qquad -  K_0\Big [ 2\pi \Big (p-1+\frac{\ell}{N} \Big ) \frac{m}{m_W} \Big ] \Big \} .
        \label{eq:lambdaell}
    \end{split}
\end{equation}
When $ m \gtrsim m_W$, this series is very well-approximated by the $p=1$ term. However, some extra work is needed to extract information about the $m\ll m_W$ case.
In Appendix \ref{sec:lambdaell}, we perform the sum over $p$ by taking the Mellin transform and find (cf. Equation \eqref{eq:r0ell}):
\begin{subequations}
    \begin{equation}
        \mathcal{R}_{0\ell}(mL)  = N\Big [\gamma + \log \frac{m}{m_W}\sin \pi \frac{\ell}{N}  +W_\ell \Big ( \frac{m}{m_W}\Big ) \Big ],
        \label{eq:well2}
    \end{equation}
where, as mentioned before, $W_\ell$ is an $O(1)$ function such that $W_\ell(0)=0$, and has a power series expansion for $\tau\equiv (m/m_W)<1$:
    \begin{equation}
            W_\ell(\tau) = \sum^\infty_{n=1}\frac{(2n)!}{(n!)^2} \Big (\frac{i \tau}{2}\Big)^{2n} \Big [ \zeta(2n+1) - Re \Big (\mathrm{Li}_{2n+1} e^{2\pi i \frac{\ell}{N}}  \Big) \Big ]\,,\qquad \tau<1 .
    \end{equation}
    \label{eq:r0ellsolution}
\end{subequations}
This is exactly Equation \eqref{eq:well}, and Equation \eqref{eq:mediumWell} follows directly from Equations \eqref{eq:lambdaell} and \eqref{eq:well2}.
Putting everything together, we finally obtain Equation \eqref{eq:kappaell}:
\begin{equation}
    \begin{split}
        \kappa_\ell = \frac{m_W^{-1}}{16\pi}  \Bigg [ \frac{8\pi^2}{Ng^2(m_We^{-\gamma})}+  b_0  \log  \frac{1}{\sin \pi \frac{\ell}{N}} 
        +\frac{2}{3} \sum_I^{n_f} W_\ell \Big (\frac{m_I}{m_W} \Big) \Bigg ]\,, \quad
         1\leq\ell\leq N-1 .
        \label{eq:kell}
    \end{split}
\end{equation}
\par
Note that although heretofore the fermion masses only appeared in the combination $mL$, Equations \eqref{eq:well2}, \eqref{eq:kell} suggest that they are in fact more naturally measured in units of $m_W$, as we should expect.
\par
On the other hand, the $44$ parts of the integrals also give us $M_{ab}^2$, the scalar (mass)$^2$ matrix. Omitting the intermediate steps,
    \begin{equation}
        \begin{split}
           M^2_{ab}   &=g^2 \sum_\beta \sum_{n=1}^\infty \frac{\beta^a\beta^b}{4\pi^2 L^2}\Big [ \sum_I^{n_f} (m_I L)^2 K_2(n m_I L) -\frac{2}{n^2}\Big ]\cos(n \beta \cdot \phi) ,
           \label{M2_result}
        \end{split}
    \end{equation}
where $K_2$ is the modified Bessel function of order 2; this matches the result from taking the second derivative of the GPY potential, \eqref{eq:potential},
which serves as a "sanity check" on our calculations. For completeness, we present $M^2_\ell$, the physical scalar (masses)$^2$:
\begin{subequations}
        \label{eq:Mell}
    \begin{equation}
        \begin{split}
           M^2_\ell &=  g^2 N m_W^2 \Bigg [ \sum_I^{n_f} F_\ell \Big  ( \frac{m_I}{m_W} \Big ) - F_\ell (0)\Bigg ],
        \end{split}
    \end{equation}
where
    \begin{equation}
        \begin{split}
            F_\ell(\tau) \equiv
         \frac{\tau^2}{4\pi^2} \sum_{p=1}^\infty \Big \{ & K_2\Big [ 2\pi\Big (p-1+\frac{\ell}{N}\Big )  \tau \Big ] +  K_2\Big [ 2\pi\Big (p-\frac{\ell}{N}\Big ) \tau \Big ] - 2 K_2( 2\pi p \tau ) \Big \} .
        \end{split}
    \end{equation}
    \label{eq:feqn}
\end{subequations}
We also present $\rho_\ell$, the eigenvalues of $\rho_{ab}$,
\begin{subequations}
    \begin{equation}
        \rho_\ell = \kappa_\ell + \frac{m_W^{-1}}{96\pi} \Bigg [ 1-\sum_I^{n_f} X_\ell \Big ( \frac{m_I}{m_W}\Big ) \Bigg ] ,
    \end{equation}
where $X_\ell$ is an $O(1)$ function defined in terms of $W_\ell$:
    \begin{equation}
        X_\ell (\tau)  = 1+ 4 \tau^2 \frac{d}{d \tau^2} W_\ell (\tau) .
    \end{equation}
    \label{eq:rhoell}
\end{subequations}
Equations \eqref{eq:Mell} and \eqref{eq:rhoell} are only presented for completeness, although they may be found without too much difficulty using the methods described in this paper.
\section{Future directions}
In this study we have derived an explicit one-loop expression for the eigenvalues of $\kappa_{ab}$, the polarisation operator of the $SU(N)$ dYM theory with massive fermions, and provisionally surveyed some properties of the emergent fourth dimension. It would be interesting to numerically examine the effect of these one-loop corrections on the $k$-string tensions, (as was done for SYM in Ref. \cite{doublestring},) but to do so would require us to compute the matrix determinants in the monopole measure, $\zeta$ --- a daunting task (see the discussion in footnote \eqref{fn:fugacity}).
\par
Additionally, the topological angle $\theta$-dependence in Yang-Mills theory has been the subject of much attention \cite{unsal12,anber13,ttt17,bonati18,bonati19}: we should also like to examine the dependence of the $k$-string tensions on the topological angle $\theta$ as well as on the circle length $L$, at the tree-order level, to compare against results on the lattice. 
\par
Finally, we would also like to further study the confining properties of dYM outside of the calculable regime, $NL\Lambda\gg1$ and its conjectured continuity with the small $NL\Lambda$ regime, on the lattice. 
\appendix
\section{The Mellin transform, and some results}
\label{sec:mellin}
In this Appendix, we explicitly evaluate the sums over $p$ in \eqref{eq:lambdaell} to obtain an expression for $\kappa_\ell$ in terms of analytic functions.
To this end, we introduce the Mellin transform, an integral transform on real-valued functions.
\begin{definition}[]
    The Mellin transform $\mathcal{M}$ is an integral transform defined on the space of real integrable functions $f:\mathbb{R}^+\to \mathbb{R}$ as:
    \begin{subequations}
    	\begin{equation}
    	    \begin{split}
    	       	\varphi(s) &\equiv \mathcal{M}_s \,[f(t) ] \\
    	       	&\equiv\int^\infty_0  dt \, t^{s-1} f(t) .
    	    \end{split}
    		\label{mellin_def}
    	\end{equation}
    In particular, for each $\lambda >0$,
        \begin{equation}
            \mathcal{M}_s[f(\lambda t)] = \lambda^{-s} \mathcal{M}_s[f(t)] .
            \label{eq:mellin_dilation}
        \end{equation}
    The inverse transform $\mathcal{M}^{-1}$ is, formally,
    	\begin{equation}
    	    \begin{split}
                f(t) &=\mathcal{M}_t^{-1}\,[\varphi(s)]\\
                &=  \frac{1}{2\pi i}\int^{c+i\infty}_{c-i\infty} ds \, t^{-s} \varphi(s) ,
    	    \end{split}
        \label{inverse_mellin}
    	\end{equation}
    \end{subequations}
    where $c$ is some real number chosen so that the integral in \eqref{inverse_mellin} converges \cite[\href{https://dlmf.nist.gov/2.5}{Sec. 2.5}]{NIST:DLMF}. Usually what this means is to take the sum over the residues of the poles of $\varphi(s)$ on the real half-line, $s\in (-\infty,c]$. To illustrate with a simple example, let us compute the Mellin transform of $f(t)=e^{-t}$, and its inverse:
\end{definition}
\begin{example}[]
\label{ex:exponential}
    Directly from the definition,
    \begin{equation}
        \begin{split}
             \mathcal{M}_s[e^{-t}] &\equiv \int^\infty_0 dt \,  t^{s-1} e^{-t}\\
                &= \Gamma(s) .
        \end{split}
    \end{equation}
    \par
    Now consider the inverse transform. Since $\Gamma(s)$ has poles at $s=0,-1,-2...$, we evaluate the integral by limiting the integration contour $c\to 0^+$ and closing the contour over the $Re(s)<0$ half-plane. The integral over the arc goes to zero at large radius, so
    \begin{equation}
        \begin{split}
          \lim_{c\to 0^+} \frac{1}{2\pi i} \int^{c+i\infty}_{c-i\infty} ds \, t^{-s} \Gamma(s) &= \sum_{n=0}^\infty \underset{s}{\text{res}}\Big ( t^{-s} \Gamma (s), -n \Big ) \\
          &\!\!\overset{\ref{gamma_series}}{=} \frac{(-t)^n}{n!}\\
          &= e^{-t} ,
        \end{split}
    \end{equation}
    as expected, because near the poles of $\Gamma(s)$,
    \begin{equation}
         \Gamma(s) = \frac{(-1)^n}{n!}\Big [ (s+n)^{-1} + \psi^{(0)}(1+n)\Big ] + O  (s+n) \,,\qquad n=0,1,2 ...
         \label{gamma_series}
    \end{equation}
    where $\psi^{(0)}(z)\equiv \frac{d}{dz} \log(\Gamma(z)) $ is the polygamma function (of order $0$).
\end{example}
\subsection{Proof of Equation \eqref{eq:logr0}}
\label{sec:r0}
We are now prepared to prove Equation \eqref{eq:logr0} and derive a series expression for $\tilde{R}_0^b$. The idea is to perform the sums over $n$ in "Mellin space," then transform back to "mass space" to obtain a series expansion in $t$. Like in Example \ref{ex:exponential}, the inverse transform involves evaluating the residue of a chain of poles on the real axis.
\par
To begin, we observe the Mellin transform of the modified Bessel function of order $\nu$, $K_\nu$, is known to be \cite[\href{http://functions.wolfram.com/03.04.22.0004.01}{03.04.22.0004.01}]{weisstein}:
\begin{equation}
	\mathcal{M}_s[K_\nu(t)] =2^{s-2}\Gamma \Big( \frac{s+\nu}{2}\Big)\Gamma\Big(\frac{s-\nu}{2}\Big) .
	\label{eq:mk0}
\end{equation}
Plugging this into Equation \eqref{R0def},
\par 
\begin{equation}
	\begin{split}
		\mathcal{M}_s[R^b_0(t)] &=  \sum_{n=1}^\infty \mathcal{M}_s [K_0\big ( n t \big)] \cos (n b) \\
		&\overset{\ref{eq:mellin_dilation}}{=}2^{s-3} \Gamma\Big(\frac{s}{2}\Big)^2 \sum_{n=1}^\infty n^{-s} ( e^{inb} +e^{-inb} ) \\
		&\overset{\ref{polylog_def}}{=} 2^{s-3}\Gamma\Big(\frac{s}{2}\Big)^2 \Big ( \textrm{Li}_s e^{i b} + \textrm{Li}_s e^{-ib} \Big ) ,
	\end{split} 
\end{equation}
where $\mathrm{Li}_{s}$ is the polylogarithm function of order $s$:
\begin{equation}
    \sum_{k=1}^\infty \frac{e^{ikb}}{k^s}=\textrm{Li}_s e^{ib} \,, \quad \text{$b$ real, $s>0$} .
    \label{polylog_def}
\end{equation}
Changing back to the original variable $t$, 
\begin{equation}
	\begin{split}
		R^b_0(t) &= \mathcal{M}_t^{-1}\mathcal{M}_s[R^b_0(t')] \\
		&= \frac{1}{2\pi i}\int^{+i\infty +c}_{-i\infty +c} \!ds\, t^{-s} 2^{s-3}\Gamma\Big(\frac{s}{2}\Big)^2  \Big (\textrm{Li}_s e^{i b} + \textrm{Li}_s e^{-ib} \Big ) . \\
	\end{split}
	\label{R0_integral}
\end{equation}
Note that the integral in Equation \eqref{R0_integral} is over the order $s$ of the polylogarithm, rather than its argument. As in Example \eqref{ex:exponential}, we can evaluate this integral by letting the integration contour approach the imaginary axis from the right, $c\to 0^+$, and close the contour over the half plane $R e(s)\leq 0$. The polylogarithm terms are regular for all $s$ for real $0<b <2\pi$, so we are left with the residues from the chain of poles at $s = 0,-2,-4...$ where the gamma function diverges.
\par 
Unfortunately, the poles of $\Gamma(s/2)^2$ are of order 2, so evaluating the residues with the integrand of \eqref{R0_integral}, as is, would involve the expression $\frac{d}{ds}\mathrm{Li}_s e^{ib}$, which produces a result that is even more opaque than our original expression.
\par 
However, a known identity \cite[\href{http://dlmf.nist.gov/25.13.3}{Eq. 25.13.3}]{NIST:DLMF} relates the polylogarithms to the Hurwitz zeta function, $\zeta$:
\par
\begin{equation}
    i^{-s}\mathrm{Li}_{s}(e^{ib})+ i^{s} \mathrm{Li}_{s}(e^{-ib}) =\frac{(2\pi )^s}{\Gamma(s)} \zeta \Big (1-s,\frac{b}{2\pi} \Big ) \,,\qquad 0<b <2\pi .
    \label{hurwitz}
\end{equation}
Where $\zeta(z,x)$ satisfies:
\begin{equation}
     \zeta  (z,x ) = \sum_{n=1}^\infty ( n+x)^{-z}\,,\quad \text{$Re(s)>1$ and $x\neq 0,1,2...$}
    \label{hurwitz_def}
\end{equation}
\par
We can sum the expression in Equation \eqref{hurwitz} with $b \to 2 \pi -b$ and divide by $(i^{s}+i^{-s})$ to rewrite the integrand of \eqref{R0_integral} as:
\begin{equation}
    \begin{split}
        t^{-s} \mathcal{M}_s[R^b_0(t')] &\overset{\ref{R0_integral}}{=} 2^{-3}\,\Gamma \Big (\frac{s}{2}\Big )^2 \Big ( \mathrm{Li}_s(e^{ib})+\mathrm{Li}_{s}(e^{-ib}) \Big ) \Big( \frac{t}{2}\Big)^{-s} \\
        &\overset{\ref{hurwitz}}{=}  \frac{2^{-4}}{\,(1+i^{-2s})} \frac{ \Gamma (\frac{s}{2} )^2}{\Gamma(s)} \Big( \frac{it}{4\pi}\Big)^{-s} \Big [ \,\zeta \Big (1-s,\frac{b}{2\pi} \Big ) + \zeta \Big (1-s,1-\frac{b}{2\pi} \Big ) \,\Big ]
    \end{split}
    \label{R0_integrand}
\end{equation}
This is helpful because the factor of $\Gamma(s)^{-1}$ in Equation \eqref{hurwitz} reduces the order of the poles by one, and the zetas in the parentheses in Equation \eqref{R0_integrand} are regular except at $s=0$, so the poles at $s=-2,-4,-6...$ are simple.
\subsubsection*{The residue at $s= 0$:}
Near $s=0$, the zeta terms diverge like $1/s$:
\begin{equation}
     \zeta(1-s,\frac{b}{2\pi}) + \zeta \Big (1-s,1-\frac{b}{2\pi} \Big )= -\frac{2}{s} -  \psi^{(0)}\!\Big (\frac{b}{2\pi}\Big )-\psi^{(0)}\!\Big (1-\frac{b}{2\pi}\Big ) + O(s) .
     \label{zeta_pole}
\end{equation}
So we must also look at the series expansion of the regular terms in \eqref{R0_integrand} near $s=0$:
\begin{equation}
  (1+i^{-2s})^{-1} \Big(\frac{i t}{4\pi}\Big )^{-s}  = \frac{1}{2}-\frac{1}{2}\log \frac{t}{4\pi}s + O(s^2) .
  \label{log_pole}
\end{equation}
Combining Equations \eqref{zeta_pole}, \eqref{log_pole} and \eqref{gamma_series}, we find our famous logarithmic term:
\begin{equation}
		\underset{s}{\mathrm{res}}  ( t^{-s} \mathcal{M}_s[R^b_0]\,,\, 0  ) = \frac{1}{2}\log \frac{t}{4\pi} - \frac{1}{4}\Big [ \, \psi^{(0)}\!\Big (\frac{b}{2\pi}\Big )+\psi^{(0)}\!\Big (1-\frac{b}{2\pi}\Big ) \, \Big  ] .
	\label{residue2}
\end{equation}
\subsubsection*{The residues at $s= -2,\,-4,\,-6...$:}
Since the poles at $s=-2,-4,-6 ...$ can only contribute terms $\sim t^{2n}$ for $n=1,2,3...$, we have proven our claim in \eqref{eq:logr0}, so we are actually \textit{done}, but since we have already done most of the work,
\begin{equation}
    \begin{split}
        \underset{s}{\mathrm{res}}  (\,  t^{-s} \mathcal{M}_s[R^b_0]\,,\,-2n \, )=& \frac{1}{4}\frac{(2n)!}{(n!)^2} \Big( \frac{it}{4\pi}\Big)^{2n}  \Big [ \,\zeta \Big (1+2n,\frac{b}{2\pi} \Big ) + \zeta \Big (1+2n,1-\frac{b}{2\pi} \Big ) \,\Big ]\\
        \overset{\ref{polygamma}}{=}&\frac{-1}{4(n!)^2} \Big( \frac{it}{4\pi}\Big)^{2n} \Big [\, \psi^{(2n)}\!\Big (\frac{b}{2\pi}\Big )+\psi^{(2n)}\!\Big (1-\frac{b}{2\pi}\Big )  \,\Big ] \,,\quad n=1,2,3... 
    \end{split}
    \label{residue1}
\end{equation}
where $\psi^{(2n)}$, the polygamma function of order $2n$, is related to $\zeta$ by \cite[\href{https://dlmf.nist.gov/25.11.12}{25.11.12}]{NIST:DLMF}
\begin{equation}
		\psi^{(2n)}(z) = -(2n)! \zeta ( 2n+1,z)\,,\qquad n=1,2,3 ...
	\label{polygamma}
\end{equation}
Putting our results together,
\begin{align}
    		\tilde{R}^b_0(t)  &= - \frac{1}{4}\sum_{n=0}^{\infty}\frac{(-1)^n}{(n!)^2} \Big  ( \frac{t}{4 \pi }\Big  )^{2n}  \Big [ \, \psi^{(2n)}\!\Big (\frac{b}{2\pi}\Big )+\psi^{(2n)}\!\Big (1-\frac{b}{2\pi}\Big ) \, \Big  ] , \label{eq:r0tilde}
\end{align}
Plugging this result into Equation \eqref{eq:kab} and taking the massless limit, $t=0$, the correction to the photon coupling matches that of the SYM result derived in Ref. \cite{symallgroups}.
\subsection{Derivation of Equation \eqref{eq:r0ellsolution}}
\label{sec:lambdaell}
Now, consider our expression for $\mathcal{R}_{0\ell}$, the eigenvalues of $\mathcal{R}_0^{ab}$, Equation \eqref{eq:lambdaell}. Let us define
\begin{subequations}
    \begin{align}
           \nu_\ell(\tau) &\equiv  \sum_{p=1}^{\infty}  K_0\Big [\Big (\frac{\ell}{N}+p-1 \Big )\tau \Big ]+ K_0 \Big [\Big(-\frac{\ell}{N}+p \Big)\tau \Big ] , \\
          \xi(\tau)&\equiv \sum_{p=1}^{\infty} 2 K_0 (p\tau) , \\
          \tau &\equiv 2\pi \frac{m}{m_W} = NLm ,
    \end{align}
\end{subequations}
Starting with $\nu_\ell$: the intermediate steps are largely the same as in the preceding subsection,
\begingroup
\begin{align*}
\allowdisplaybreaks
        \nu_\ell(\tau)\equiv &\, \mathcal{M}_t^{-1}\mathcal{M}_s \Bigg \{ \sum_{p=1}^\infty  K_0\Big [ \tau\Big ( p-1+\frac{\ell}{N}\Big )\Big ] + K_0\Big [\tau\Big ( p-\frac{\ell}{N} \Big )\Big ] \Bigg \} \\ 
        \overset{\ref{eq:mellin_dilation}}{=}
        &\,\mathcal{M}_t^{-1} \Big [\, \tau^{-s} 2^{s-2} \Gamma\Big(\frac{s}{2}\Big)^2 \sum_{p=1}^{\infty} \Big[ \Big ( p-1+\frac{\ell}{N} \Big )^{-s}+ \Big ( p-\frac{\ell}{N} \Big )^{-s} \Big ]\\
        \overset{\ref{hurwitz_def}}{=} 
        &\,\sum_{\text{poles}} \underset{s}{\text{res}}\Big[ \,\tau^{-s} 2^{s-2} \Gamma\Big(\frac{s}{2}\Big)^2  \Big [ \zeta\Big ( s,1-\frac{\ell}{N} \Big )+\zeta\Big ( s,\frac{\ell}{N} \Big )\Big ]\,,\, \{ \text{pole} \} \, \Big ]\\
       =
          & \,\frac{\pi}{\tau} + \sum_{n=0}^\infty  \frac{1}{(n!)^2}\Big(\frac{\tau}{2}\Big )^{2n} \Big[ \zeta'\Big ( -2n,1-\frac{\ell}{N} \Big ) + \zeta'\Big ( -2n,\frac{\ell}{N} \Big ) \\ &\qquad\qquad+ \Big[ \cancelto{0}{\zeta \Big (  -2n,1-\frac{\ell}{N} \Big ) +  \zeta \Big (  -2n,\frac{\ell}{N} \Big ) } \Big ] \Big (\psi(n+1)-\log\frac{\tau}{2} \Big ) \Big] \\
        \overset{\ref{bernoulli}}{=}
          &\,\frac{\pi}{\tau} + \sum_{n=0}^\infty  \frac{1}{(n!)^2}\Big(\frac{\tau}{2}\Big )^{2n} \Big[ \zeta'\Big ( -2n,1-\frac{\ell}{N} \Big ) + \zeta'\Big ( -2n,\frac{\ell}{N} \Big ) \Big] \\
        \overset{\ref{nice_identity}}{=}
        &\,\frac{\pi}{\tau} +\frac{1}{2} \sum_{n=0}^\infty  \frac{(2n)!}{(n!)^2} \Big(\frac{i \tau}{4 \pi }\Big )^{2n} \Big( \mathrm{Li}_{2n+1} e^{2\pi i \frac{\ell}{N}} + \mathrm{Li}_{2n+1} e^{-2\pi i \frac{\ell}{N}} \Big ) ,
\end{align*}
\endgroup
The expression in the third line has order-$2$ poles at $s=0,-2,-4...$ (from the gammas,) and a simple pole at $s=1$, (from the zetas,) and $\zeta'(s,x) \equiv \frac{d}{ds}\zeta(s,x) $.
\par 
The striked out term in the fourth line vanishes because \cite[\href{https://dlmf.nist.gov/25.11.14}{25.11.14}]{NIST:DLMF} 
\begin{equation}
        \zeta(-2n,x) = -\frac{B_{2n+1}(x)}{2n+1} = \frac{B_{2n+1}(-x)}{2n+1} \,,\qquad x\,\,\text{real}\,,\, n= 0, 1, 2 \, ...\\
    \label{bernoulli}
\end{equation}
where $B_{k}(x)$ is the Bernoulli polynomial of order $k$, which has parity $(-1)^{k}$ under $x\to 1-x$.
\par 
Lastly, the final line follows from \cite[Eq. 13]{adamchik97}:
\begin{equation}
    \begin{split}
        \zeta'(-2n,1-x) + \zeta'(-2n,x) &= \frac{(2n)!}{(2\pi i)^{2n}} \Big (  \mathrm{Li}_{2n+1}e^{2\pi i x} + \mathrm{Li}_{2n+1}e^{-2\pi i x}\Big ) , \\
        & \qquad \qquad \qquad \qquad\qquad x\,\,\text{real}\,,\, n= 0, 1, 2 \, ...\\
    \end{split}
    \label{nice_identity}
\end{equation}
and since $\mathrm{Li}_{1}(z)= -\log (1-z)$,
\begin{equation}
   \begin{split}
        \nu_\ell(\tau)=\frac{\pi}{\tau}& +\log \frac{1}{2\sin \pi \frac{\ell}{N}} \\ &+\frac{1}{2} \sum_{n=1}^\infty \frac{(2n)!}{(n!)^2} \Big(\frac{i\tau}{4 \pi }\Big )^{2n} \Big ( \mathrm{Li}_{2n+1} e^{2\pi i \frac{\ell}{N}} + \mathrm{Li}_{2n+1} e^{-2\pi i \frac{\ell}{N}} \Big ) .
   \end{split}
\end{equation}
To solve for $\xi$, we observe that, for the plain (Riemann) zetas $\zeta(z,0)\equiv \zeta(z)$ \cite[\href{http://functions.wolfram.com/10.01.20.0022.01}{10.01.20.0022.01}]{weisstein},
\begin{equation}
        \zeta'(-2n) =\frac{(2n)!}{(2\pi i)^n} \zeta(2n+1) , \\
\end{equation}
and the sum over $p$ goes through almost verbatim. The result is:
\begingroup
\begin{equation}
    \begin{split}
          \xi(\tau)&=\,\sum_{\text{poles}} \underset{s}{\text{res}}\Big [\,\tau^{-s} 2^{s-2} \Gamma\Big(\frac{s}{2}\Big)^2  \zeta(s)\,,\, \{ \text{pole} \} \, \Big ]\\
          &=\frac{\pi}{\tau} + \gamma +\log\frac{\tau}{4\pi} + \sum^\infty_{n=1}\Big (\frac{i \tau}{4\pi}\Big)^{2n}\frac{(2n)!}{(n!)^2}\zeta(2n+1) .
    \end{split}
    \label{xi}
\end{equation}
As before, the poles are located at $s=0,-2,-4....$  and $s=1$. So together,
\begin{equation}
    \begin{split}
        N^{-1}\mathcal{R}_{0\ell} &= \xi - \nu_\ell  \\
        &=\gamma + \log \Big ( \sin \pi \frac{\ell}{N}\tau \Big ) + W_\ell (\tau) ,
        \label{eq:r0ell}
    \end{split}
\end{equation}
where $W_\ell$ has a power series expansion in $\tau$:
\begin{subequations}
    \begin{equation}
        W_\ell(\tau) \equiv \sum^\infty_{n=1}\frac{(2n)!}{(n!)^2} \Big (\frac{i \tau}{2}\Big)^{2n} \Big [ \zeta(2n+1) -\frac{1}{2}\Big (\mathrm{Li}_{2n+1} e^{2\pi i \frac{\ell}{N}} + \mathrm{Li}_{2n+1} e^{-2\pi i \frac{\ell}{N}} \Big) \Big ] .
\end{equation}
As mentioned in Footnote \ref{fn:roottest}, the root test shows that this infinite series diverges for $m\geq m_W$. In that case, $W_\ell$ can be approximated by
\begin{equation}
    \begin{split}
        W_\ell(\tau) &= \gamma + \log  \sin \pi \frac{\ell}{N} \tau  + 2K_0(  2\pi  \tau )  \\
        & \quad - K_0 \Big [ 2\pi  \Big ( 1- \frac{\ell}{N} \Big )\tau\Big ] -  K_0\Big ( 2\pi \frac{\ell}{N} \tau \Big ) + O(e^{-2 \pi\tau }) .
    \end{split}
\end{equation}
\par
Manipulating the series expansions for $\zeta(2n+1)$ and $\mathrm{Li}_{2n+1}$, Equations \eqref{polylog_def} and \eqref{hurwitz_def}, we can also write $W_\ell(\tau)$ purely in terms of elementary functions:
\begin{equation}
        W_\ell(\tau) =  2\sum_{k=1}^{\infty}  \Big [ \Big (k^2+\tau^2 \Big )^{-1/2} -k^{-1}\Big ] \sin^2 \Big ( \pi\frac{\ell}{N} k\Big ) .
    \label{eq:wellnegative}
\end{equation}
\end{subequations}
We note that Equation \eqref{eq:wellnegative} converges much more slowly than the previous ones, but on the other hand, it clearly shows that $W_\ell$ is strictly negative, and goes to zero as $\frac{\ell}{N} \to 0$. Observing that 
\begin{equation}
    \frac{d^2}{dx^2} \mathrm{Li}_{2n+1}( e^{i x} ) =  - \mathrm{Li}_{2n-1}( e^{i x} ) ,
\end{equation}
a similar argument shows that $W_\ell$ is strictly concave up in $\ell$ for all $1\leq\ell\leq \floor{N/2}$.
\par
Finally, Equation \eqref{eq:lambdaell} shows that $\mathcal{R}_{0\ell}\to 0$ rapidly as $\tau \to \infty$, so \eqref{eq:r0ell} demands 
\begin{equation}
    W_\ell(\tau) \to  \log \frac{e^{-\gamma}}{\tau \sin \pi \frac{\ell}{N}}\,,\quad \tau\to \infty ,
\end{equation}
which concludes the proofs for the statements we made about $W_\ell$ in Equation \eqref{eq:wellproperties}.
\section*{Acknowledgements}
The author thanks E. Poppitz for many helpful discussions throughout the production of this work. This work is funded by an NSERC Discovery Grant.
\medskip

\providecommand{\href}[2]{#2}\begingroup\raggedright\endgroup

\end{document}